\pdfoutput=1
\documentclass[11pt,letterpaper]{article}

\usepackage[margin=1in]{geometry}
\usepackage[utf8]{inputenc}
\usepackage[T1]{fontenc}
\usepackage{lmodern}
\usepackage{microtype}
\usepackage{setspace}
\usepackage{amsmath,amssymb,amsfonts,amsthm,mathtools}
\usepackage{bm}
\usepackage{graphicx}
\usepackage{array,booktabs,multirow,tabularx}
\usepackage{pifont}
\usepackage{tikz}
\usetikzlibrary{positioning,arrows.meta,fit,backgrounds,calc}
\usepackage{xcolor}
\usepackage[authoryear,round]{natbib}
\usepackage[
  colorlinks=true,
  linkcolor=blue!60!black,
  citecolor=blue!60!black,
  urlcolor=blue!60!black,
  pdftitle={Sparse Multivariate Distribution-to-Distribution Regression via Dependent Mixtures -- A Bayesian Semiparametric Approach},
  pdfauthor={Jintao Wen, Yang Ni, and Weixuan Zhu}
]{hyperref}
\usepackage[nameinlink,capitalize]{cleveref}

\numberwithin{equation}{section}
\allowdisplaybreaks

\newtheorem{proposition}{Proposition}
\theoremstyle{definition}

\newtheorem{remark}{Remark}

\crefname{proposition}{proposition}{propositions}
\Crefname{proposition}{Proposition}{Propositions}
\crefname{remark}{remark}{remarks}
\Crefname{remark}{Remark}{Remarks}

\newcommand{\cmark}{\ding{51}}
\newcommand{\xmark}{\ding{55}}

\newcommand{\spider}{SPIDER}

\title{\Large\bfseries Sparse Multivariate Distribution-to-Distribution Regression via Dependent Mixtures -- A Bayesian Semiparametric Approach}

\author{%
  Jintao Wen$^{1}$, \quad
  Yang Ni$^{3,*}$, \quad
  and Weixuan Zhu$^{1,2,*}$\\[0.8em]
  \footnotesize\itshape $^{1}$Department of Statistics and Data Science, School of Economics, Xiamen University, Xiamen, China\\
  \footnotesize\itshape $^{2}$Wang Yanan Institute for Studies in Economics, Xiamen University, Xiamen, China\\
  \footnotesize\itshape $^{3}$Department of Statistics and Data Sciences, The University of Texas at Austin, Austin, TX, USA\\[0.45em]
  \footnotesize\normalfont Email: \texttt{wenjintao1@gmail.com}\\
  \footnotesize\normalfont $^{*}$Corresponding authors: \texttt{yang.ni@austin.utexas.edu}, \texttt{zhuweixuan@gmail.com}
}

\date{}

\begin{document}

\maketitle

\begin{abstract}
We propose a Bayesian semiparametric model for sparse multivariate distribution-to-distribution regression that accommodates multiple distributional predictors and responses of arbitrary and potentially differing dimensions and performs model selection. Both predictor and response distributions are represented as finite mixtures with subject-specific mixing weights and shared component parameters across subjects. Bijective transformations map mixing weights from the simplex to unconstrained Euclidean spaces, turning the distributional regression into a multivariate linear model on latent coordinates. A group spike-and-slab prior on the regression coefficients enables variable selection at the level of distributional predictors, yielding posterior inclusion probabilities with natural uncertainty quantification. We develop an MCMC scheme that combines P\'olya--Gamma augmentation for the predictor coordinates with collapsed--uncollapsed sampling for the regression layer, using a Laplace approximation for the collapsed response-coordinate update. The remaining blocks are updated directly from their conditional distributions, avoiding Metropolis--Hastings tuning. Unlike many optimal transport-based methods, the framework operates directly on empirical distributions without requiring continuity assumptions. Simulations demonstrate competitive predictive accuracy and robust variable selection under both well-specified and misspecified settings. Applied to population-scale single-cell RNA sequencing data from the OneK1K study, the method infers sparse cell-cell communication networks by identifying which sender cell-type expression distributions drive variation in receiver expression.

\medskip
\noindent\textbf{Keywords:} Bayesian variable selection; Cell-cell communication; Finite mixture model; Single-cell RNA sequencing; Spike-and-slab prior.
\end{abstract}

\section{Introduction}
\label{sec:introduction}

In the classical regression framework, both predictors and responses are scalars or finite-dimensional vectors. However, modern data collection increasingly produces observations that are naturally represented as probability distributions. In genomics, single-cell sequencing yields expression distributions across cell populations; in economics, income distributions characterize the economy of regions; in environmental science, pollutant concentrations follow complex distributional patterns. When distributional variables serve as predictors while the response remains scalar, the problem of density regression arises \citep{poczos2013distribution, szabo2016learning, matabuena2023distributional}. Conversely, when real-valued predictors are used to model a distributional response, one obtains density-on-scalar regression \citep*{dunson2007bayesian}. More recently, the distribution-to-distribution regression (DDR) problem, where both predictors and responses are distributional variables, has attracted growing attention \citep*{oliva2013distribution}.

A growing body of literature addresses DDR under various settings. For univariate distributions with a single predictor, \citet{ghodrati2022distribution} represent the regression operator as a monotone optimal transport map, \citet*{chen2023wasserstein} develop Wasserstein regression using parallel transport, and \citet{zhao2023density} propose density-on-density regression via the Fisher--Rao metric. When multiple univariate predictors are available, \citet*{kou2019compact} propose a distribution regression network that encodes distributions as discretized histograms, \citet{pegoraro2022projected} develop projected Wasserstein methods that represent distributions via quantile function projections onto B-spline bases, \citet{chen2024distribution} introduce distribution-in-distribution-out (DIDO) regression using parallel transport, and \citet{chen2026multi} develop multi-transport regression with weighted Fr\'echet means. In the multivariate setting, \citet{zhu2026geodesic} develop geodesic optimal transport (GOT) regression accommodating multiple multivariate predictors via a transport algebra. Recently, \citet*{nguyen2025bayesian,nguyen2026bayesian} propose Bayesian frameworks for DDR based on sliced Wasserstein for single and multiple multivariate distributional predictors. Despite this progress, existing methods still have limitations. Most approaches rely on optimal transport or Riemannian geometry, which require the underlying distributions to be continuous to access the Riemannian structure and guarantee the existence of transport maps \citep*{nguyen2026bayesian}; they cannot be applied directly to empirical distributions arising from finite samples. Furthermore, for high-dimensional regression, variable selection is routinely performed in the classic setting but has not been rigorously investigated in distributional regression. Table~\ref{tab:comparison} summarizes these gaps.

\begin{table}[t!]
\centering
\caption{Comparison of DDR methods. Columns indicate whether a method supports: multivariate distributions (Multiv.), multiple distributional predictors (Multiple), discrete empirical distributions (Disc.), flexible dimensionality between predictor and response (Flex.\ Dim.), variable selection (VS).}
\label{tab:comparison}
\begingroup
\small
\setlength{\tabcolsep}{6pt}
\renewcommand{\arraystretch}{1.02}
\begin{tabular}{@{}lccccc@{}}
\toprule
Method & Multiv. & Multiple & Disc. & Flex.\ Dim. & VS \\
\midrule
\citet*{chen2023wasserstein}   & \xmark & \xmark & \xmark & \xmark & \xmark \\
\citet{zhao2023density}       & \xmark & \xmark & \xmark & \xmark & \xmark \\
\citet*{kou2019compact}        & \xmark & \cmark & \xmark & \xmark & \xmark \\
\citet{pegoraro2022projected} & \xmark & \cmark & \xmark & \xmark & \xmark \\
\citet{chen2024distribution}  & \xmark & \cmark & \xmark & \xmark & \xmark \\
\citet{chen2026multi}         & \xmark & \cmark & \xmark & \xmark & \xmark \\
\citet{zhu2026geodesic}       & \cmark & \cmark & \xmark & \xmark & \xmark \\
\citet*{nguyen2025bayesian}    & \cmark & \xmark & \cmark & \xmark & \xmark \\
\citet*{nguyen2026bayesian}    & \cmark & \cmark & \cmark & \cmark & \xmark \\
\spider                       & \cmark & \cmark & \cmark & \cmark & \cmark \\
\bottomrule
\end{tabular}
\endgroup
\end{table}

To enable variable selection and retain all the desirable properties of existing methods, we propose \spider{}
(\textbf{SP}arse d\textbf{I}stribution-to-\textbf{D}istribution
r\textbf{E}gression via mixtu\textbf{R}es), a new Bayesian method for sparse multivariate distribution-to-distribution regression, which
\begin{enumerate}
    \item models both predictor and response distributions as finite mixtures, accommodating arbitrary and potentially differing dimensions;
    \item maps mixing weights from the probability simplex to unconstrained Euclidean coordinates through computationally motivated bijections, thereby recasting distribution-to-distribution regression as multivariate linear regression on latent coordinates;
    \item incorporates group spike-and-slab priors for variable selection at the distributional predictor level;
    \item develops an MCMC scheme that combines P\'olya--Gamma augmentation for the predictor coordinates with collapsed--uncollapsed sampling for the regression layer, using a Laplace approximation for the collapsed response-coordinate update and direct conditional updates for the remaining blocks, thereby avoiding Metropolis--Hastings tuning; and
    \item operates directly on empirical distributions from finite samples, without requiring the continuity assumptions needed by many existing methods based on optimal transport or Riemannian geometry.
\end{enumerate}
In many practical settings, predictor and response distributions describe fundamentally different physical entities occupying distinct feature spaces, making equal-dimensionality assumptions unrealistic. The proposed framework is broadly applicable; we demonstrate its utility on single-cell RNA sequencing data for inferring cell-cell communication networks, where identifying which signaling distributions drive response variation is a primary scientific goal \citep{yazar2022single, wilk2024comparative}.

The remainder of the article is organized as follows. Section~\ref{sec:model} introduces the hierarchical mixture model framework and the distributional regression formulation. Section~\ref{sec:inference} develops the posterior inference procedure, including specific transformation choices, the MCMC sampler, and the posterior predictive distribution. Section~\ref{sec:simulations} presents simulation studies. Section~\ref{sec:application} applies the method to single-cell RNA sequencing data. Section~\ref{sec:discussion} concludes with a discussion. Proofs, detailed posterior-computation derivations, additional simulation results, application details, and MCMC diagnostics are provided in Appendices~\ref{supp:proof_identifiability}--\ref{supp:mcmc_diagnostics}.

\section{Model}
\label{sec:model}

We consider data collected from $N$ subjects, indexed by $i = 1, \ldots, N$. For each subject $i$, the distribution of a population of interest serves as the response, denoted by $\mathbf{Y}_i$, while the distributions of $G$ distinct populations serve as the predictors, denoted by $\mathbf{X}_i^{(1)}, \ldots, \mathbf{X}_i^{(G)}$. \spider{} models these distributional variables through a hierarchical Bayesian finite mixture model, linking them via a latent regression with variable selection.

\subsection{Mixture Representations}
\label{sec:mixture}

We represent both response and predictor distributions as finite mixtures with shared component parameters across subjects and subject-specific mixing proportions. For the response, each subject $i$ is associated with a $d_Y$-variate distribution $\mathbf{Y}_i$ from which we observe a sample of size $N_i^Y$, denoted $\mathcal{Y}_i = \{\mathbf{y}_{i,c}\}_{c=1}^{N_i^Y} \subset \mathbb{R}^{d_Y}$, where $c$ indexes individual observations within subject $i$ and $\mathbf{y}_{i,c} \sim \mathbf{Y}_i$. We model $\mathbf{Y}_i$ as a mixture of $K$ components:
\begin{equation*}
p(\mathbf{y}_{i,c} \mid \bm{\pi}_i^Y, \bm{\Theta}^Y) = \sum_{k=1}^K \pi_{i,k}^Y \, \mathcal{K}_Y(\mathbf{y}_{i,c} \mid \bm{\theta}_k^Y),
\end{equation*}
where $\bm{\Theta}^Y = \{\bm{\theta}_k^Y\}_{k=1}^K$ are population-shared component parameters, $\bm{\pi}_i^Y \in \Delta^{K-1}$ are subject-specific mixing weights, and $\mathcal{K}_Y$ is a mixture kernel appropriate for the data type. For continuous data in $\mathbb{R}^{d_Y}$, the Gaussian kernel $\mathcal{K}_Y(\cdot \mid \bm{\theta}_k^Y) = \mathcal{N}(\cdot \mid \bm{\mu}_k^Y, \bm{\Sigma}_k^Y)$ is a natural default; for heavy-tailed data one may use a Student-$t$ kernel, for counts a Poisson kernel, and for categorical outcomes a multinomial kernel.

Similarly, for each predictor $g \in \{1, \ldots, G\}$, subject $i$ has a $d_g$-variate predictor distribution $\mathbf{X}_i^{(g)}$ from which we observe a sample of size $N_i^{(g)}$, denoted $\mathcal{X}_i^{(g)} = \{\mathbf{x}_{i,c}^{(g)}\}_{c=1}^{N_i^{(g)}} \subset \mathbb{R}^{d_g}$, with $\mathbf{x}_{i,c}^{(g)} \sim \mathbf{X}_i^{(g)}$. Each predictor distribution is modeled as a mixture of $J_g$ components:
\begin{equation*}
p(\mathbf{x}_{i,c}^{(g)} \mid \bm{\pi}_i^{(g)}, \bm{\Theta}^{(g)}) = \sum_{j=1}^{J_g} \pi_{i,j}^{(g)} \, \mathcal{K}_g(\mathbf{x}_{i,c}^{(g)} \mid \bm{\theta}_{g,j}^X),
\end{equation*}
where $\bm{\Theta}^{(g)} = \{\bm{\theta}_{g,j}^X\}_{j=1}^{J_g}$ are shared parameters and $\bm{\pi}_i^{(g)} \in \Delta^{J_g-1}$ captures subject-specific heterogeneity. The dimensions $d_Y$ and $d_g$ may differ across variables, and the within-subject sample sizes $N_i^Y$ and $N_i^{(g)}$ may vary across subjects.

The assumption that component parameters $\bm{\Theta}^Y$ and $\bm{\Theta}^{(g)}$ are shared across subjects, while mixing weights vary, is motivated by three considerations. First, shared kernels define a common ``dictionary'' of distributional features, enabling meaningful cross-subject comparisons: differences between subjects are expressed entirely through the compositional vector $\bm{\pi}_i$, which admits natural geometric structure on the simplex. Second, sharing component parameters provides statistical borrowing of strength across subjects, which is particularly beneficial when within-subject sample sizes are small. Third, the regression in our framework targets the mixing weights specifically; if component parameters also varied across subjects, the regression signal would be confounded with changes in the mixture basis, complicating both inference and interpretation.

\begin{proposition}[Identifiability up to a common permutation]
\label{prop:mixture_identifiability}
For fixed \(L\), let
\(f_i(\mathbf{x})
= \allowbreak \sum_{\ell=1}^{L}\pi_{i\ell}
\phi_d(\mathbf{x};\bm{\mu}_\ell,\bm{\Sigma}_\ell)\),
\(i=1,\ldots,N\), and define \(\widetilde f_i\) analogously using
\((\widetilde\pi_{i\ell},
\widetilde{\bm\mu}_\ell,\widetilde{\bm\Sigma}_\ell)\).
Assume that all covariance matrices are positive definite, the
component parameter pairs are pairwise distinct within each component
dictionary, and
\(\bm\pi_i,\widetilde{\bm\pi}_i
\in\operatorname{int}(\Delta^{L-1})\).
If \(f_i=\widetilde f_i\) almost everywhere for every \(i\), then
there exists a permutation \(\sigma\in S_L\), common to all subjects
in the block, such that
\[
(\widetilde{\bm\mu}_{\sigma(\ell)},
 \widetilde{\bm\Sigma}_{\sigma(\ell)})
=(\bm\mu_\ell,\bm\Sigma_\ell),
\qquad
\widetilde\pi_{i,\sigma(\ell)}=\pi_{i\ell},
\]
for all \(i=1,\ldots,N\) and \(\ell=1,\ldots,L\).
\end{proposition}

The proof is provided in Appendix~\ref{supp:proof_identifiability}. Once a labeling convention is fixed, the bijectivity of \(h_Y\) and \(h_X^{(g)}\) makes the latent coordinates identifiable from the mixing weights.

\subsection{Distributional Regression via Latent Coordinates}
\label{sec:d2d_regression}

The primary objective is to model the functional dependence of the response distribution on the predictor distributions through their mixing weights: $(\bm{\pi}_i^{(1)}, \ldots, \bm{\pi}_i^{(G)}) \mapsto \bm{\pi}_i^Y$. Since mixing weights lie on the probability simplex $\Delta$, we map them to unconstrained Euclidean spaces via bijective transformations.

Let \(h_Y:\operatorname{int}(\Delta^{K-1})\to\mathbb R^{K-1}\) denote a bijective transformation for the response mixing weights, and let \(h_X^{(g)}:\operatorname{int}(\Delta^{J_g-1})\to \mathbb R^{J_g-1}\) denote analogous transformations for each predictor $g$. These transformations share two key properties: (i) they are bijections between the open simplex and the full Euclidean space $\mathbb{R}^{d-1}$ and (ii) they enable efficient Bayesian inference in the transformed space, which we exploit in Section~\ref{sec:inference}. Define the latent response coordinates $\mathbf{U}_i = h_Y(\bm{\pi}_i^Y) \in \mathbb{R}^{K-1}$ and the latent predictor coordinates $\mathbf{V}_i^{(g)} = h_X^{(g)}(\bm{\pi}_i^{(g)}) \in \mathbb{R}^{J_g-1}$ for each predictor $g$. The distributional dependency is modeled by linking these latent coordinates through a multivariate linear regression (although it can be extended to nonlinear settings with e.g., basis expansion):
\begin{equation*}
\mathbf{U}_i
=
\boldsymbol{\alpha}
+
\mathbf{B}^\top \mathbf{V}_i
+
\bm{\varepsilon}_i,
\qquad
\bm{\varepsilon}_i \sim \mathcal{N}(\mathbf{0}, \bm{\Sigma}_U),
\end{equation*}
where $\mathbf{V}_i = [(\mathbf{V}_i^{(1)})^\top, \ldots, (\mathbf{V}_i^{(G)})^\top]^\top \in \mathbb{R}^Q$ with $Q = \sum_{g=1}^G (J_g - 1)$ is formed by concatenating all predictor coordinates, $\mathbf{B} \in \mathbb{R}^{Q \times (K-1)}$ is the regression coefficient matrix, $\boldsymbol{\alpha}\in\mathbb{R}^{K-1}$ is an intercept vector, and $\bm{\Sigma}_U$ is the residual covariance.

\begin{remark}
\label{rem:full_identifiability}
After fixing a labeling convention within each mixture block, Proposition~\ref{prop:mixture_identifiability} and the bijectivity of \(h_Y\) and \(h_X^{(g)}\) uniquely determine the subject-specific mixing weights and latent coordinates. If each \(\bm\Sigma_V^{(g)}\) is positive definite, the stacked predictor coordinate \(\mathbf V_i\) has full-dimensional support in \(\mathbb R^Q\). Consequently, the conditional Gaussian law
\[
\mathbf U_i\mid\mathbf V_i=\mathbf v \sim \mathcal N\!\left( \boldsymbol\alpha+\mathbf B^\top\mathbf v, \bm\Sigma_U \right)
\]
uniquely determines \((\boldsymbol\alpha,\mathbf B,\bm\Sigma_U)\). Thus, all continuous structural parameters are identifiable after the mixture labels are fixed.
\end{remark}

The specific transformation choices, selected to enable efficient posterior computation, are deferred to Section~\ref{sec:inference}.

\subsection{Prior Specification}
\label{sec:priors}

We specify priors for each layer of the hierarchical model, beginning with the standard conjugate choices for the mixture parameters and working up to the central inferential component of our framework: the group spike-and-slab prior for variable selection.

\paragraph{Mixture component priors.}
For the remainder of the paper we specialize the mixture kernels to the Gaussian family, setting $\mathcal{K}_Y(\cdot \mid \bm{\theta}_k^Y) = \mathcal{N}(\cdot \mid \bm{\mu}_k^Y, \bm{\Sigma}_k^Y)$ and $\mathcal{K}_g(\cdot \mid \bm{\theta}_{g,j}^X) = \mathcal{N}(\cdot \mid \bm{\mu}_{g,j}^X, \bm{\Sigma}_{g,j}^X)$. We assign independent Normal-Inverse-Wishart (NIW) priors for the component parameters. For the response:
\begin{equation*}
    (\bm\mu^Y_k, \bm\Sigma^Y_k) \sim \mathcal{NIW}(\bm\xi_Y, \kappa_Y, \nu_Y, \bm\Psi_Y), \quad k=1, \ldots, K,
\end{equation*}
and analogously for each predictor block $g = 1, \ldots, G$:
\begin{equation*}
    (\bm\mu^X_{g,j}, \bm\Sigma^X_{g,j}) \sim \mathcal{NIW}(\bm\xi_g, \kappa_g, \nu_g, \bm\Psi_g), \quad j = 1, \ldots, J_g.
\end{equation*}
These are standard choices that permit closed-form Gibbs updates.

\paragraph{Latent predictor coordinate prior.}
To complete the hierarchy, we place a Gaussian prior on the latent predictor coordinates $\mathbf{V}_i^{(g)}$, with a conjugate Normal-Inverse-Wishart hyperprior on their mean and covariance:
\begin{equation*}
    \mathbf{V}_i^{(g)} \mid \bm{\mu}_V^{(g)}, \bm{\Sigma}_V^{(g)} \sim \mathcal{N}\!\left(\bm{\mu}_V^{(g)}, \bm{\Sigma}_V^{(g)}\right),
\end{equation*}
\begin{equation*}
    \bigl(\bm{\mu}_V^{(g)}, \bm{\Sigma}_V^{(g)}\bigr) \sim \mathcal{NIW}\!\left(\mathbf{m}_{V0}^{(g)}, \lambda_{V0}^{(g)}, \nu_{V0}^{(g)}, \bm{\Psi}_{V0}^{(g)}\right).
\end{equation*}
This prior encodes the belief that the latent coordinates for each predictor are exchangeable across subjects, and supports conjugate posterior updates for the population-level parameters $(\bm{\mu}_V^{(g)}, \bm{\Sigma}_V^{(g)})$.

\paragraph{Regression-layer prior and variable selection.}
Following \citet*{brown1998multivariate}, we decompose the regression-layer prior as
\[
\pi(\boldsymbol{\alpha}, \mathbf{B}, \bm\Sigma_U, \bm\gamma)
=
\pi(\boldsymbol{\alpha}\mid\bm\Sigma_U)\,
\pi(\mathbf{B}\mid\bm\Sigma_U, \bm\gamma)\,
\pi(\bm\gamma)\,
\pi(\bm\Sigma_U),
\]
where $\bm\gamma = (\gamma_1, \ldots, \gamma_G)^\top \in \{0,1\}^G$ is a latent binary vector of predictor inclusion indicators for variable selection, specified in detail below.

We assign the intercept a vague prior
\begin{equation*}
    \boldsymbol{\alpha} \mid \bm\Sigma_U \sim \mathcal{N}(\boldsymbol{\alpha}_0,\, h_\alpha\bm\Sigma_U), \qquad h_\alpha \to \infty.
\end{equation*}
Conditional on $\bm\gamma$, we place a conjugate prior on the active regression rows and the residual covariance:
\begin{align*}
    \mathbf{B}_{(\bm\gamma)}
    \mid \bm\Sigma_U,\bm\gamma
    &\sim
    \mathcal{MN}\!\left(
        \mathbf{0},\,
        \mathbf{H}_{(\bm\gamma)},\,
        \bm\Sigma_U
    \right),\\
    \bm\Sigma_U
    &\sim
    \mathcal{IW}_q(\nu_U,\bm\Psi_U),
\end{align*}
where $q=K-1$, $\mathbf{B}_{(\bm\gamma)}$ is the $p_\gamma\times (K-1)$ submatrix of active rows, with $p_\gamma=\sum_{g:\gamma_g=1}(J_g-1)$. Rows corresponding to $\gamma_g=0$ are fixed at zero under the selection prior. The active row covariance is
\begin{equation*}
    \mathbf{H}_{(\bm\gamma)}
    =
    \mathbf{D}_{(\bm\gamma)}
    \mathbf{R}_{(\bm\gamma)}
    \mathbf{D}_{(\bm\gamma)} .
\end{equation*}
We set $\mathbf{R}_{(\bm\gamma)}=\mathbf I_{p_\gamma}$ and $\mathbf{D}_{(\bm\gamma)}=\tau_B\mathbf I_{p_\gamma}$, so that $\mathbf{H}_{(\bm\gamma)}=\tau_B^2\mathbf I_{p_\gamma}$.

The variable selection is driven by the latent binary vector $\bm\gamma$ with independent Bernoulli prior,
\begin{equation*}
    \gamma_g \sim \mathrm{Bernoulli}(w_g), \quad g = 1, \ldots, G.
\end{equation*}
Each predictor $g$ is represented in the regression through $J_g - 1$ latent coordinates $\mathbf{V}_i^{(g)}$, corresponding to $J_g - 1$ consecutive rows of the coefficient matrix $\mathbf{B}$. Since these rows collectively parameterize a single distributional variable, including some while excluding others would have no obvious interpretation. Therefore, the group indicator $\gamma_g$ acts jointly on all $J_g - 1$ rows: when $\gamma_g = 0$, every corresponding row of $\mathbf{B}$ is set to zero; when $\gamma_g = 1$, these rows are estimated from the data.

\begin{figure}[htbp]
\centering
\resizebox{0.92\linewidth}{!}{%
\begin{tikzpicture}[
    >=Stealth,
    node distance=1.2cm and 1.4cm,
    latent/.style={circle, draw, minimum size=22pt, inner sep=0pt, font=\small},
    obs/.style={circle, draw, fill=gray!25, minimum size=22pt, inner sep=0pt, font=\small},
    param/.style={circle, draw, minimum size=22pt, inner sep=0pt, font=\small},
    hyper/.style={rectangle, draw, rounded corners=2pt, minimum size=18pt, inner sep=2pt, font=\small},
    plate/.style={draw, thick, rounded corners=3pt, inner sep=10pt},
    platelabel/.style={font=\footnotesize, anchor=south east, inner sep=2pt},
    every edge/.style={draw, ->, >=Stealth, semithick},
]

\node[hyper] (muV) {$\bm{\mu}_V^{(g)}\!,\bm{\Sigma}_V^{(g)}$};

\node[latent, below=1.0cm of muV] (V) {$\mathbf{V}_i^{(g)}$};

\node[param, left=1.6cm of V] (thetaX) {$\bm{\theta}_{g,j}^X$};

\node[obs, below=1.0cm of V] (X) {$\mathbf{x}_{i,c}^{(g)}$};

\draw[->] (muV) -- (V);
\draw[->] (thetaX) -- (X);
\draw[->] (V) -- (X);

\node[plate, fit=(X), label={[platelabel]south east:{$N_i^{(g)}$}}] (plateC) {};

\node[plate, fit=(V)(plateC), yshift=-3pt, xshift=3pt, label={[platelabel]south east:{$N$}}] (plateI_X) {};

\node[plate, fit=(thetaX), label={[platelabel]south east:{$J_g$}}] (plateJ) {};

\node[plate, dashed, fit=(muV)(plateI_X)(plateJ), inner sep=14pt, label={[platelabel]south east:{$G$}}] (plateG) {};

\node[latent, right=3.8cm of V] (U) {$\mathbf{U}_i$};

\node[latent, above=0.7cm of U, xshift=-1.0cm] (gamma) {$\bm{\gamma}$};

\node[hyper, above=0.7cm of U, xshift=1.0cm] (BSigma) {$\boldsymbol\alpha,\mathbf{B},\bm{\Sigma}_U$};

\node[param, right=1.6cm of U] (thetaY) {$\bm{\theta}_{k}^Y$};

\node[obs, below=1.0cm of U] (Y) {$\mathbf{y}_{i,c}$};

\node[hyper, above=0.5cm of gamma] (w) {$w_g$};

\draw[->] (V) -- node[above, font=\scriptsize, pos=0.5] {} (U);
\draw[->] (gamma) -- (U);
\draw[->] (BSigma) -- (U);
\draw[->] (U) -- (Y);
\draw[->] (thetaY) -- (Y);
\draw[->] (w) -- (gamma);

\node[plate, fit=(Y), label={[platelabel]south east:{$N_i^Y$}}] (plateCY) {};
\node[plate, fit=(U)(plateCY), yshift=-3pt, xshift=3pt, label={[platelabel]south east:{$N$}}] (plateI_Y) {};
\node[plate, fit=(thetaY), label={[platelabel]south east:{$K$}}] (plateK) {};

\end{tikzpicture}%
}
\caption{Plate diagram of the full hierarchical model. Shaded nodes denote observed data, and the dashed plate indexes the $G$ distributional predictors. The indicators $\bm{\gamma}$ control the inclusion of predictor-coordinate blocks in the regression for $\mathbf U_i$.}
\label{fig:plate_diagram}
\end{figure}
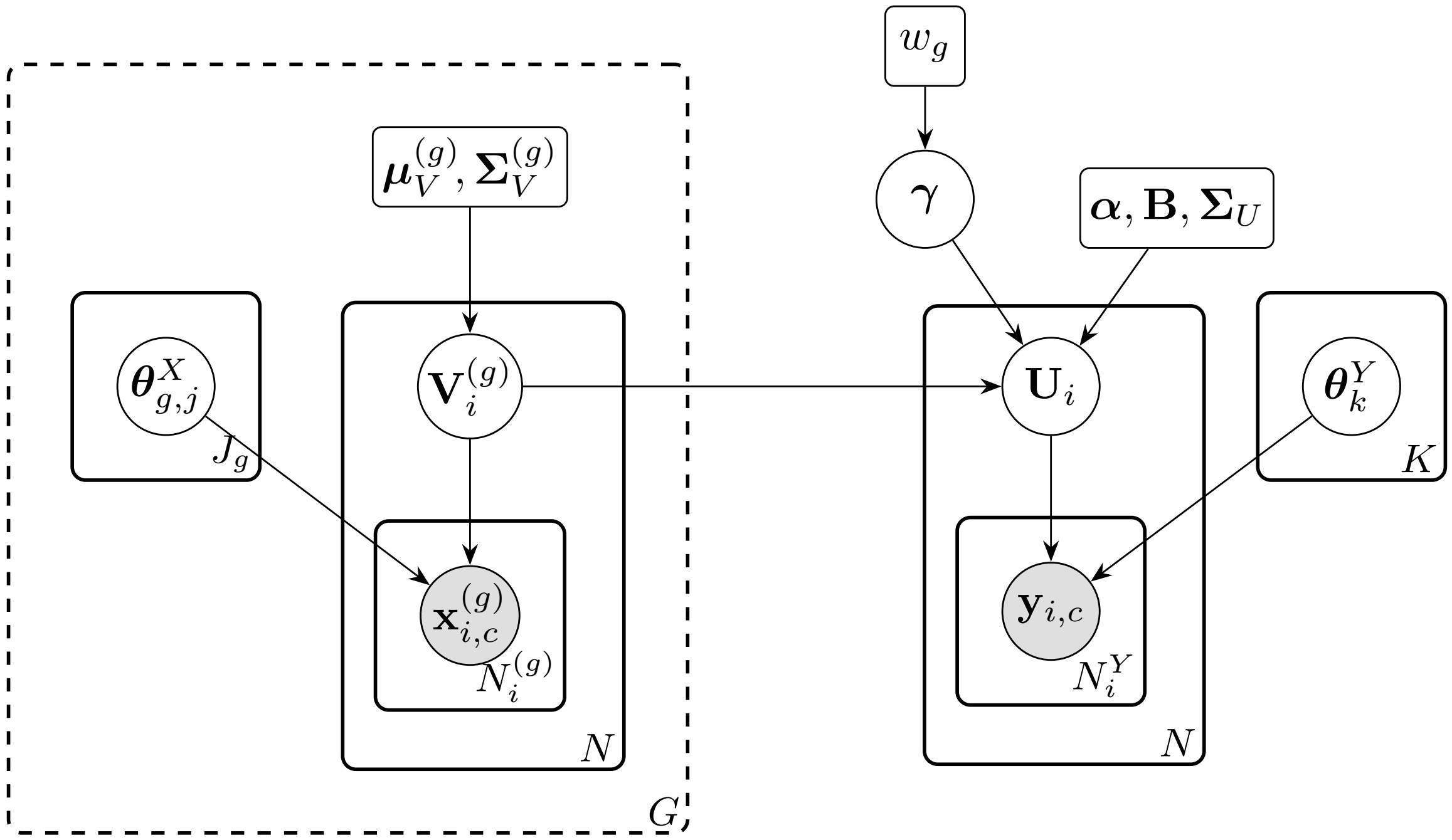

The full hierarchical model is represented by the plate diagram in \Cref{fig:plate_diagram}.

\section{Posterior Inference}
\label{sec:inference}

We develop a sampler that avoids Metropolis--Hastings steps by combining P\'olya--Gamma data augmentation with collapsed--uncollapsed sampling. This section first specifies the transformations $h_Y$ and $h_X^{(g)}$ introduced in Section~\ref{sec:d2d_regression}, and then presents the MCMC sampler and the posterior predictive distribution. Complete derivations, conditional updates, and implementation details are provided in Appendix~\ref{supp:mcmc_inference}.

\subsection{Transformation Choices}
\label{sec:transformations}

For the response mixing weights $\bm\pi^Y_i \in \Delta^{K-1}$, we employ the additive log-ratio (ALR) transformation:
\begin{equation*}
    \mathbf{U}_i = h_Y(\bm\pi^Y_i) = \mathrm{ALR}_K(\bm\pi^Y_i) = \left[ \log\left(\frac{\pi^Y_{i,1}}{\pi^Y_{i,K}}\right), \ldots, \log\left(\frac{\pi^Y_{i,K-1}}{\pi^Y_{i,K}}\right) \right]^\top \in \mathbb{R}^{K-1}.
\end{equation*}
Given $\mathbf V$ and $\bm\gamma$, integrating out $\boldsymbol{\alpha},\mathbf B,\bm\Sigma_U$ under the NIW prior yields a matrix-$t$ marginal prior for $\mathbf U$. Because the multinomial likelihood for the response allocation counts is not conjugate to this matrix-$t$ prior, we follow the collapsed Laplace strategy of \citet{silverman2022bayesian} to update $\mathbf U$ from an approximation to $p(\mathbf U\mid \mathbf V,\bm\gamma,\mathbf z^Y)$; more details are provided in Appendix~\ref{app:cu_sampling}.

For each predictor $g$, we adopt a logistic stick-breaking (SB) parameterization. The mixing weights $\bm{\pi}_i^{(g)}$ are transformed into latent logits $\mathbf{V}_i^{(g)} = [V_{i1}^{(g)}, \ldots, V_{i,J_g-1}^{(g)}]^\top \in \mathbb{R}^{J_g-1}$ via:
\begin{equation*}
    \pi_{i,j}^{(g)} = \sigma(V_{ij}^{(g)}) \prod_{\ell=1}^{j-1} (1 - \sigma(V_{i\ell}^{(g)})), \quad j = 1, \ldots, J_g - 1,
\end{equation*}
where $\sigma(z) = (1 + e^{-z})^{-1}$ and $\pi_{i,J_g}^{(g)} = \prod_{\ell=1}^{J_g-1}(1 - \sigma(V_{i\ell}^{(g)}))$. This parameterization is chosen because it decomposes the multinomial likelihood into a sequence of conditionally independent binomial factors, each admitting P\'olya--Gamma (PG) augmentation \citep*{polson2013bayesian, linderman2015dependent}. The augmentation introduces auxiliary variables that render the conditional posterior of $\mathbf{V}_i^{(g)}$ Gaussian, enabling block Gibbs updates without Metropolis--Hastings steps.

\subsection{MCMC Sampler}
\label{sec:posterior_computation}

The posterior algorithm alternates among four blocks: (i) mixture components and allocation indicators; (ii) P\'olya--Gamma-augmented predictor coordinates; (iii) collapsed updates of the response coordinates and inclusion indicators; and (iv) uncollapsed updates of the regression parameters.

Conditioned on the current component allocation indicators $z_{i,c}^Y$ and $z_{i,c}^{(g)}$, the Gaussian mixture component parameters admit conjugate NIW updates. For the response, label switching is resolved by sorting the components according to the first coordinate of their means at each iteration. For the predictors, identifiability is enforced via sequential truncated sampling on the first coordinate of each component mean, maintaining the strict ordering $\mu_{g,1,1}^X < \cdots < \mu_{g,J_g,1}^X$ during sampling. Separately, conditioned on the current predictor latent coordinates $\{\mathbf V_i^{(g)}\}_{i=1}^N$, the population-level predictor-coordinate parameters $(\bm{\mu}_V^{(g)}, \bm{\Sigma}_V^{(g)})$ are updated via conjugate NIW posteriors.

For the predictor latent coordinates, we use the logistic stick-breaking representation of the multinomial likelihood. Following \citet*{linderman2015dependent}, the allocation likelihood is written as a product of binomial-logit kernels. For each predictor block \(g\), subject \(i\), and stick-breaking coordinate \(j=1,\ldots,J_g-1\), we introduce
\[
    \omega_{ij}^{(g)}
    \mid
    N_{ij}^{(g)}, V_{ij}^{(g)}
    \sim
    \mathrm{PG}
    \left(
        N_{ij}^{(g)},
        V_{ij}^{(g)}
    \right).
\]
This augmentation gives a Gaussian contribution to the update of the stacked predictor coordinate \(\mathbf V_i\). The definitions of \(N_{ij}^{(g)}\), \(\bm\kappa_i\), and \(\bm\Omega_i\) are given in Appendix~\ref{app:polya_gamma}.

Let $\bm\mu_V=[(\bm\mu_V^{(1)})^\top,\ldots,(\bm\mu_V^{(G)})^\top]^\top$ and $\bm\Sigma_V=\operatorname{blockdiag}\{\bm\Sigma_V^{(1)},\ldots,\bm\Sigma_V^{(G)}\}$.

The row-wise full conditional is
\[
    \mathbf V_i
    \mid
    \bm\kappa_i,\bm\Omega_i,
    \mathbf U_i,\boldsymbol\alpha,
    \mathbf B,\bm\Sigma_U,
    \bm\mu_V,\bm\Sigma_V
    \sim
    \mathcal N_Q
    \left(
        \bm\Lambda_i^{-1}\bm\eta_i,\,
        \bm\Lambda_i^{-1}
    \right),
\]
where
\begin{align*}
\bm\Lambda_i
&=
\bm\Sigma_V^{-1}
+\bm\Omega_i
+\mathbf B\bm\Sigma_U^{-1}\mathbf B^\top,\\
\bm\eta_i
&=
\bm\Sigma_V^{-1}\bm\mu_V
+\bm\kappa_i
+\mathbf B\bm\Sigma_U^{-1}
(\mathbf U_i-\boldsymbol\alpha).
\end{align*}

Third, the response coordinates $\mathbf U$ are updated after marginalizing
$(\boldsymbol\alpha,\mathbf B_{(\bm\gamma)},\bm\Sigma_U)$ from the
regression layer. Let
$n_{i,k}^Y=\sum_c\mathbf 1\{z_{i,c}^Y=k\}$ and
$\bm\pi_i^Y=\mathrm{ALR}_K^{-1}(\mathbf U_i)$. The resulting
collapsed target is
\begin{equation}
\label{eq:collapsed_U_target}
p(\mathbf U\mid\mathbf z^Y,\mathbf V,\bm\gamma)
\propto
\prod_{i=1}^N\prod_{k=1}^K
\{\pi_{i,k}^Y(\mathbf U_i)\}^{n_{i,k}^Y}
\left|\mathbf Q_\gamma[\mathbf U]\right|^{-a_N},
\qquad
a_N=\frac{N+\nu_U}{2},
\end{equation}
where the centered residual-scale matrix \(\mathbf Q_\gamma[\mathbf U]\), together with the matrix \(\mathbf K_\gamma\) used below, is defined in Appendix~\ref{app:cu_sampling}.

Writing $\mathbf u=\operatorname{vec}(\mathbf U)$, let
$\widehat{\mathbf u}$ be the mode of
\eqref{eq:collapsed_U_target} and $\mathbf H_U$ its negative Hessian.
Following \citet{silverman2022bayesian}, we sample $\mathbf u$ from
the Gaussian Laplace approximation
$\mathcal N(\widehat{\mathbf u},\mathbf H_U^{-1})$.

The same marginalization yields the group-selection score
\begin{align*}
m(\bm\gamma)
&=
-\frac q2\log\det\mathbf H_{(\bm\gamma)}
-\frac q2\log\det\mathbf K_\gamma
-a_N\log\det\mathbf Q_\gamma,\\
\Pr(\gamma_g=1\mid\bm\gamma_{-g},\mathbf U,\mathbf V)
&=
\operatorname{logit}^{-1}\!\left[
m(\bm\gamma^{(1)})-m(\bm\gamma^{(0)})
+\log\frac{w_g}{1-w_g}
\right],
\end{align*}
where $\bm\gamma^{(b)}$ is the current inclusion vector with its
$g$th element set to $b$. Posterior inclusion probabilities are estimated by the retained-sample frequencies of
$\gamma_g=1$.

After updating $\mathbf U$ and $\bm\gamma$, the regression block is uncollapsed by drawing $\bm\Sigma_U$, $\mathbf B_{(\bm\gamma)}$, and $\boldsymbol\alpha$ from their conjugate inverse-Wishart, matrix-normal, and Gaussian conditional distributions, respectively.

Finally, the response and predictor allocations are refreshed from their categorical full conditionals using the current mixing weights and component parameters.

\subsection{Posterior Predictive Distribution}
\label{sec:prediction}

Given training data $\mathcal{D}$ and observed predictor samples $\mathcal{X}_{\mathrm{new}}^{(g)}$ for a new subject across all $G$ predictors, we predict the response distribution by Monte Carlo averaging over the $T$ retained posterior draws after burn-in.
Let
\[
\bm{\theta}^{(t)} = \bigl\{\bm{\gamma},\;
\mathbf{B},\; \bm{\Sigma}_U,\;
(\bm{\mu}_k^Y, \bm{\Sigma}_k^Y)_k,\;
(\bm{\mu}_{g,j}^X, \bm{\Sigma}_{g,j}^X)_{g,j},\;
(\bm{\mu}_V^{(g)}, \bm{\Sigma}_V^{(g)})_g
\bigr\}^{(t)}
\]
denote the sampled parameters at iteration $t$, together with the training sample means $\bar{\mathbf{U}}^{(t)} = N^{-1}\sum_{i=1}^N \mathbf{U}_i^{(t)}$ and $\bar{\mathbf{V}}^{(t)} = N^{-1}\sum_{i=1}^N \mathbf{V}_i^{(t)}$ stored at each iteration.

For each draw $t$ and each predictor $g$ with $\gamma_g^{(t)}=1$, we obtain the new subject's latent coordinates $\mathbf{V}_{\mathrm{new}}^{(g,t)}$ by first drawing $\mathbf{V}_{\mathrm{new},0}^{(g,t)} \sim \mathcal{N}(\bm{\mu}_V^{(g,t)}, \bm{\Sigma}_V^{(g,t)})$, then assigning component indicators $z_c^{(g)} \sim p(z_c^{(g)} \mid \mathbf{x}_c^{(g)}, \mathbf{V}_{\mathrm{new},0}^{(g,t)}, \bm{\Theta}^{(g,t)})$, drawing P\'olya--Gamma auxiliaries, and sampling $\mathbf{V}_{\mathrm{new}}^{(g,t)}$ from the resulting Gaussian posterior. For inactive predictors with $\gamma_g^{(t)}=0$, $\mathbf{V}_{\mathrm{new}}^{(g,t)}$ is set to zero. The results are concatenated to form $\mathbf{V}_{\mathrm{new}}^{(t)}$.

Since $\boldsymbol{\alpha}$ has been analytically marginalized
under the vague conjugate intercept prior
(Section~\ref{sec:priors}), the predictive distribution for the
new subject's response coordinate is
\begin{equation}
\label{eq:pred_U_sample}
    \mathbf{U}_{\mathrm{new}}^{(t)}
\mid \mathbf{V}_{\mathrm{new}}^{(t)},\mathcal D
\sim \mathcal{N}_q\!\left(
    \bar{\mathbf{U}}^{(t)} + (\mathbf{B}^{(t)})^\top
    \bigl(\mathbf{V}_{\mathrm{new}}^{(t)} -
    \bar{\mathbf{V}}^{(t)}\bigr),\;
    \bigl(1+\tfrac{1}{N}\bigr)
    \bm{\Sigma}_U^{(t)}\right).
\end{equation}
In practice, for each
draw $t$ we approximate
$\mathbb{E}[\mathrm{ALR}_K^{-1}(\mathbf{U}_{\mathrm{new}})
\mid \bm{\theta}^{(t)}, \mathcal{X}_{\mathrm{new}}]$
by averaging $\mathrm{ALR}_K^{-1}$ over a small number of
draws from~\eqref{eq:pred_U_sample}, yielding the estimated
mixing weights $\widehat{\bm{\pi}}_{\mathrm{new}}^{Y(t)}$.
The posterior predictive density is
\begin{equation*}
    \hat{p}(\mathbf{y}_{\mathrm{new}} \mid
    \mathcal{X}_{\mathrm{new}}, \mathcal{D}) \approx
    \frac{1}{T} \sum_{t=1}^{T} \sum_{k=1}^{K}
    \widehat{\pi}_{\mathrm{new},k}^{Y(t)}\,    \mathcal{N}\!\left(\mathbf{y}_{\mathrm{new}} \mid
    \bm{\mu}_k^{Y(t)}, \bm{\Sigma}_k^{Y(t)}\right).
\end{equation*}
To generate samples, we draw a posterior index $t$ uniformly
from $\{1, \ldots, T\}$, a component
$k \sim \mathrm{Categorical}( \widehat{\bm{\pi}}_{\mathrm{new}}^{Y(t)})$, and finally
$\mathbf{y}_{\mathrm{new}} \sim
\mathcal{N}(\bm{\mu}_k^{Y(t)}, \bm{\Sigma}_k^{Y(t)})$.

\section{Simulation Studies}
\label{sec:simulations}

We report two primary simulation scenarios. Scenario~1 evaluates predictive accuracy and variable selection under a well-specified Gaussian mixture data-generating mechanism, whereas Scenario~2 examines predictive robustness to model misspecification. Both scenarios compare \spider{} with Geodesic Optimal Transport regression \citep[GOT;][]{zhu2026geodesic}. An additional \(d=5\) experiment comparing \spider{} with the distribution regression network \citep*[DRN;][]{kou2019compact} and projected Wasserstein regression \citep[Projected Reg.;][]{pegoraro2022projected} is reported in Appendix~\ref{sec:sim_univariate}.

Throughout this section, we generate \(N=400\) subjects, divided equally into training and test sets, with two active distributional predictors. $K_0$ and $J_{g,0}$ denote the true numbers of response and predictor mixture components, respectively. We set $J_{g,0}=J_0$ for all $g$ and vary $K_0=J_0\in\{3,4,5\}$. For \spider{}, we select the fitted component counts $(K, J_g)$ via minimizing the Widely Applicable Information Criterion (WAIC). The complete data-generating mechanisms and scenario-specific parameter settings are provided in Appendix~\ref{sec:simulation_dgp}.

\subsection{Evaluation Metrics}

We evaluate predictive accuracy using the sliced Wasserstein distance (SWD), which compares multivariate distributions by averaging one-dimensional Wasserstein distances over random projections \citep{rabin2012wasserstein,bonneel2015sliced,peyre2019computational}. Specifically, for two distributions $\mu, \nu \in \mathcal{P}_p(\mathbb{R}^d)$, the Sliced Wasserstein-$p$ distance is
\begin{equation}
\label{eq:swd}
    SW_p(\mu, \nu) = \left( \int_{\mathbb{S}^{d-1}} W_p^p(\theta^\sharp \mu,\, \theta^\sharp \nu)\, \mathrm{d}\sigma(\theta) \right)^{1/p},
\end{equation}
where $\theta^\sharp \mu$ denotes the pushforward of $\mu$ under the projection $\mathbf{z} \mapsto \theta^\top \mathbf{z}$, $W_p$ is the univariate Wasserstein-$p$ distance, and $\sigma$ is the uniform measure on the unit sphere $\mathbb{S}^{d-1}$. In practice, we approximate~\eqref{eq:swd} by a Monte Carlo average over $L$ random projections drawn uniformly from $\mathbb{S}^{d-1}$, with each univariate $W_p$ computed via the sorting-based closed form. Throughout our experiments, we set $p = 2$ and $L = 1000$.

To evaluate variable selection, we calculate the true positive rate (TPR), false positive rate (FPR), and Matthews correlation coefficient (MCC), where a predictor is declared active whenever its posterior inclusion probability exceeds $0.5$.

\subsection{Simulation Results}
\label{sec:sim_summary}

Each setting is replicated 100 times. For every replication, each candidate \spider{} model is fitted using 2,000 MCMC iterations, with the first 1,000 discarded as burn-in.

\paragraph{Scenario 1: Well-specified model.}
Scenario 1 examines predictive accuracy and variable selection under a correctly specified Gaussian mixture model. We set $d=2$, $G=10$ candidate distributional predictors, two truly active predictors, $K_0=J_0\in\{3,4,5\}$, and $C\in\{50,200\}$ observations per subject. GOT is given oracle access to the two active predictors, whereas \spider{} performs variable selection over all ten candidates by thresholding the posterior inclusion probability $P(\gamma_g=1)$ at $0.5$. Results are reported in Table~\ref{tab:sim_wellspec}.

\begin{table}[htbp]
\centering
\caption{Scenario 1 (well specified): predictive accuracy and variable selection.}
\label{tab:sim_wellspec}
\begingroup
\small
\setlength{\tabcolsep}{6pt}
\renewcommand{\arraystretch}{1.02}
\begin{tabular}{@{}ccccccc@{}}
\toprule
& & \multicolumn{2}{c}{SWD (SE)} & \multicolumn{3}{c}{\spider{} Variable Selection} \\
\cmidrule(lr){3-4} \cmidrule(lr){5-7}
$C$ & $K_0=J_0$ & \spider{} & GOT & MCC & TPR & FPR \\
\midrule
50  & 3 & \textbf{0.665} (0.006) & 0.703 (0.005) & 0.876 & 0.985 & 0.064 \\
50  & 4 & \textbf{0.962} (0.007) & 1.011 (0.007) & 0.945 & 1.000 & 0.031 \\
50  & 5 & \textbf{1.271} (0.008) & 1.314 (0.008) & 0.925 & 1.000 & 0.042 \\
200 & 3 & \textbf{0.647} (0.006) & 0.701 (0.006) & 0.957 & 0.995 & 0.021 \\
200 & 4 & \textbf{0.915} (0.007) & 1.009 (0.007) & 0.995 & 1.000 & 0.002 \\
200 & 5 & \textbf{1.216} (0.009) & 1.311 (0.008) & 0.995 & 1.000 & 0.002 \\
\bottomrule
\end{tabular}
\endgroup
\end{table}

Across all settings, \spider{} achieves lower SWD than GOT. The relative reductions range from $3.3\%$ to $5.4\%$ for $C=50$ and from $7.2\%$ to $9.3\%$ for $C=200$. Variable selection is stable, with MCC at least $0.876$, TPR at least $0.985$, and FPR at most $0.064$.

\paragraph{Scenario 2: Misspecified model.}
Scenario 2 examines robustness to model misspecification by replacing all Gaussian mixture components in the data-generating mechanism with multivariate $t$ components while fitting \spider{} with Gaussian mixture components. Specifically, observations for both response and predictor variables are drawn from $t_\nu(\bm\mu_k,\bm\Sigma_k)$ rather than $\mathcal N(\bm\mu_k,\bm\Sigma_k)$. We focus on the strong misspecification regime $\nu=3$, set $d=2$ and $C=50$, and vary $K_0=J_0\in\{3,4,5\}$. Both \spider{} and GOT are given oracle access to the active predictor set so that the comparison focuses on predictive robustness. Results are reported in Table~\ref{tab:sim_misspec}.

\begin{table}[htbp]
\centering
\caption{Scenario 2 (misspecified): predictive performance under $t_3$-mixture data.}
\label{tab:sim_misspec}
\begingroup
\small
\setlength{\tabcolsep}{7pt}
\renewcommand{\arraystretch}{1.02}
\begin{tabular}{@{}ccc@{}}
\toprule
$K_0=J_0$ & \spider{} SWD (SE) & GOT SWD (SE) \\
\midrule
3 & \textbf{0.779} (0.009) & 0.844 (0.005) \\
4 & \textbf{1.013} (0.006) & 1.095 (0.006) \\
5 & \textbf{1.279} (0.008) & 1.340 (0.008) \\
\bottomrule
\end{tabular}
\endgroup
\end{table}

Under strong misspecification, \spider{} consistently achieves lower SWD than GOT, with relative improvements ranging from $4.6\%$ to $7.7\%$. Thus, the Gaussian finite-mixture approximation remains effective even when the true component distributions are substantially heavier tailed.

Overall, \spider{} yields the lowest SWD in all scenarios and maintains high TPR with low FPR in the variable-selection settings.

\section{Application: Immune Cell Communication in Human Peripheral Blood}
\label{sec:application}

We apply \spider{} to infer cell-cell communication (CCC) from population-scale single-cell RNA sequencing (scRNA-seq) data. As surveyed by \citet{dimitrov2022comparison}, standard CCC inference methods rely on agglomerative strategies that collapse single-cell measurements into cluster-level means to estimate interaction scores between cell types. However, \citet{wilk2024comparative} demonstrate that such aggregation can cause the loss of at least $50\%$ of ground-truth communication states by failing to capture fine-grained interactions. Because CCC operates between individual cells rather than groups, aggregating expression values masks important heterogeneity and specificity. Biological signals in scRNA-seq data are often encoded in the complex shape of the expression distribution (e.g., multimodality, differential proportions among subpopulations), not merely in global mean shifts \citep{korthauer2016statistical}.

By modeling expression profiles as full probability distributions rather than summary statistics, \spider{} preserves the variability and heterogeneity essential for uncovering specific communication pathways, particularly those driven by rare but highly active cell subsets.

\subsection{Data and Problem Setup}
\label{sec:application_setup}

We analyze scRNA-seq data from the OneK1K study \citep{yazar2022single}. The data contain five major immune lineages: B cells, CD4$^+$ T cells, CD8$^+$ T cells, NK cells, and monocytes. After filtering to individuals with at least 50 cells per cell type, we retain 200 subjects for training and 23 for held-out evaluation. We use CellChat \citep{jin2021inference} to identify significant ligand-receptor (L-R) interactions with $p < 0.05$. This yields 153 significant L-R pairs across all cell type combinations. The ligand and receptor genes used for each target cell type are summarized in Appendix~\ref{supp:application_details}.

Formally, for each receiver cell type in individual $i$, the receptor expression vectors $\mathcal{Y}_i \subset \mathbb{R}^{d_Y}$ serve as the response, and for each sender cell type $g$, the ligand expression vectors $\mathcal{X}_i^{(g)} \subset \mathbb{R}^{d_g}$ serve as predictors; both are modeled via the mixture framework of Section~\ref{sec:model}. All expression vectors are standardized per gene across all individuals.

We fit \spider{} for each of the five cell types as a target. Model complexity $(K, J_g)$ is selected via WAIC over a grid $\{3, 4, 5\}$, with the MCMC sampler run for $10{,}000$ iterations ($5{,}000$ burn-in). MCMC convergence is verified via trace plots and conditional autocorrelation functions, conditioning on the sender-inclusion pattern when the chain visits multiple $\bm{\gamma}$ states (Figures~\ref{fig:mcmc_diag_cd4}--\ref{fig:mcmc_diag_mono} in Appendix~\ref{supp:mcmc_diagnostics}). We do not include GOT as a baseline because it requires predictor and response distributions to share the same dimension, whereas ligands and receptors can have vastly different dimensions (Table~\ref{tab:supp_LR_genes} in Appendix~\ref{supp:application_details}). As a baseline, we fit a multivariate group lasso on mean-summarized distributions using the \texttt{adelie} package \citep{yang2024fast}: for each sender $g$ and subject $i$, the ligand and receptor distributions are reduced to their sample means $\bar{\mathbf{x}}_i^{(g)}$ and $\bar{\mathbf{y}}_i$, and a group lasso is fitted from the stacked mean predictors to $\bar{\mathbf{y}}_i$ with sender groups penalized jointly. To evaluate the mean-based group lasso under the same distributional metric, we converted its held-out mean prediction into an empirical distribution. For held-out subject $i$, the fitted group lasso gives a predicted receptor mean $\widehat{\bm\mu}_i$. We estimated a common residual covariance $\widehat{\bm\Sigma}_{\mathrm{res}}$ from the training residuals and generated $n_{\mathrm{pred}}$ pseudo-cells from $\mathcal N(\widehat{\bm\mu}_i,\widehat{\bm\Sigma}_{\mathrm{res}})$. The resulting empirical distribution was compared with the held-out response sample using SWD.

\subsection{Results}
\label{sec:application_results}

Table~\ref{tab:real_data_results} summarizes the selected senders and predictive performance for each target cell type.

\begin{table}[htbp]
\centering
\caption{Inference results for cell-cell communication.}
\label{tab:real_data_results}
\begingroup
\small
\setlength{\tabcolsep}{6pt}
\renewcommand{\arraystretch}{1.02}
\begin{tabular}{@{}llcc@{}}
\toprule
\textbf{Target} & \textbf{Method} & \textbf{Selected Senders} & \textbf{SWD (Mean $\pm$ SE)} \\
\midrule
CD4$^+$ T & \spider        & CD8$^+$ T, NK, Monocytes & $0.184 \pm 0.013$ \\
          & Group Lasso & CD8$^+$ T                & $0.795 \pm 0.009$ \\
\midrule
CD8$^+$ T & \spider        & B, CD4$^+$ T, NK         & $0.264 \pm 0.015$ \\
          & Group Lasso & CD4$^+$ T                & $0.878 \pm 0.017$ \\
\midrule
NK        & \spider        & CD4$^+$ T, B, CD8$^+$ T, Monocytes & $0.303 \pm 0.022$ \\
          & Group Lasso & ---                       & $1.045 \pm 0.071$ \\
\midrule
B         & \spider        & CD4$^+$ T, NK, CD8$^+$ T & $0.374 \pm 0.019$ \\
          & Group Lasso & NK                        & $1.152 \pm 0.026$ \\
\midrule
Monocytes & \spider        & NK, CD8$^+$ T, CD4$^+$ T & $0.419 \pm 0.025$ \\
          & Group Lasso & ---                       & $1.318 \pm 0.050$ \\
\bottomrule
\end{tabular}
\endgroup
\end{table}

In terms of predictive performance, \spider{} substantially outperforms the mean-based group lasso across all targets (Table~\ref{tab:real_data_results}), with SWD reductions ranging from approximately $67.5\%$ to $76.9\%$. The group lasso fails to identify any active senders for the NK and monocyte targets, and only one sender for the remaining targets. This is biologically implausible and illustrates a key limitation of the mean-based approaches: in single-cell data, communication signals are often encoded in the shape of the distribution, rather than in a global shift of the mean. \spider{} recovers these distributional signals through explicit mixture modeling, demonstrating enhanced sensitivity to dependencies invisible to first-order statistics.

\begin{figure}[htbp]
\centering
\begin{tikzpicture}[
    scale=1.05,
    cell/.style={
        draw,
        rounded corners=2pt,
        thick,
        align=center,
        inner xsep=4pt,
        inner ysep=2pt,
        font=\small
    },
    recip/.style={
        <->,
        >=stealth,
        thick,
        shorten <=4pt,
        shorten >=4pt
    },
    directed/.style={
        ->,
        >=stealth,
        thick,
        densely dashed,
        shorten <=4pt,
        shorten >=4pt
    }
]

\node[cell] (NK)   at (0,2.2)       {NK};
\node[cell] (CD4)  at (-2.55,0.65)  {CD4$^+$ T};
\node[cell] (CD8)  at (2.55,0.65)   {CD8$^+$ T};
\node[cell] (Mono) at (-1.55,-1.55) {Monocytes};
\node[cell] (B)    at (1.55,-1.55)  {B};

\draw[recip] (CD4) to[bend left=10]  (CD8);
\draw[recip] (CD4) to[bend left=10]  (NK);
\draw[recip] (CD8) to[bend right=10] (NK);
\draw[recip] (CD4) to[bend right=10] (Mono);
\draw[recip] (CD8) to[bend left=10]  (B);
\draw[recip] (NK)  to[bend left=12]  (B);
\draw[recip] (NK)  to[bend right=12] (Mono);

\draw[directed] (CD4) to[bend left=8]  (B);
\draw[directed] (CD8) to[bend right=8] (Mono);

\draw[recip] (-1.9,-2.45) -- (-1.3,-2.45);
\node[anchor=west,font=\footnotesize]
    at (-1.2,-2.45) {reciprocal selection};

\draw[directed] (-1.9,-2.85) -- (-1.3,-2.85);
\node[anchor=west,font=\footnotesize]
    at (-1.2,-2.85) {one-way selection};

\end{tikzpicture}
\caption{Inferred cell-cell communication network from the OneK1K scRNA-seq data.}
\label{fig:ccc_network}
\end{figure}
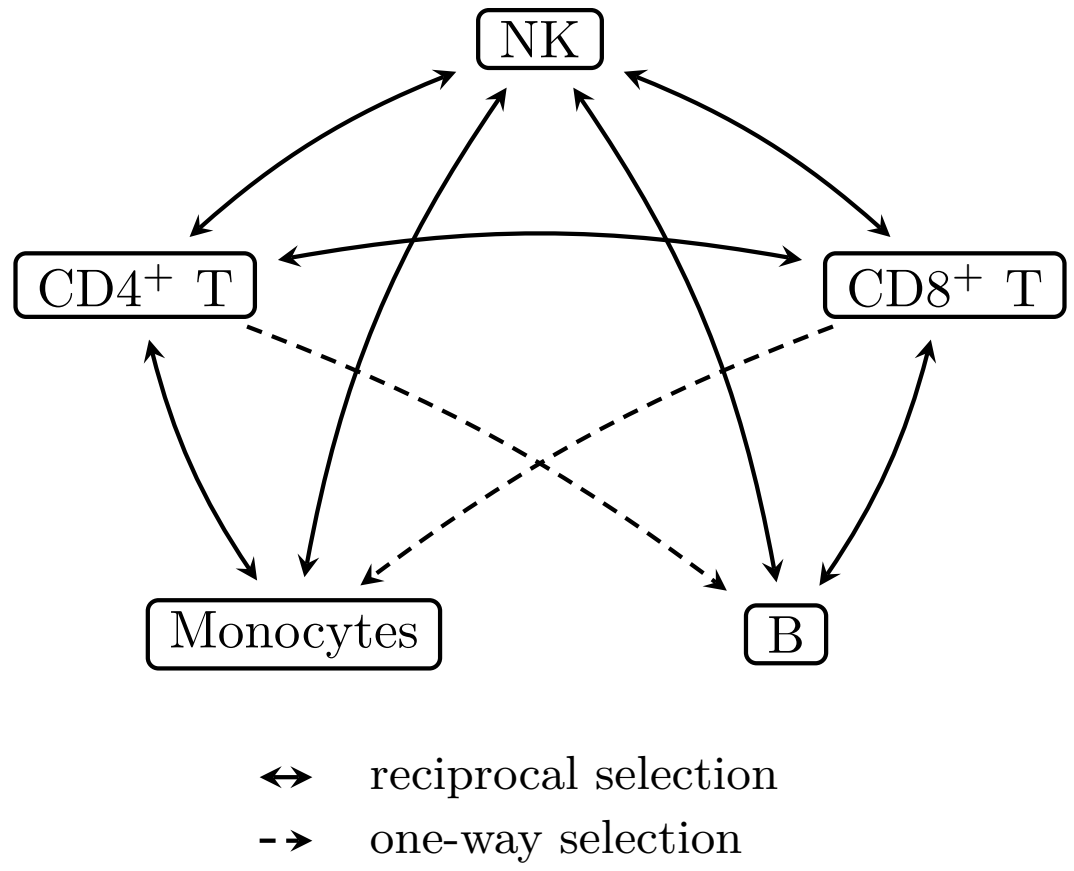

The inferred network in \Cref{fig:ccc_network} is considerably denser than the mean-based baseline and reveals broad lymphoid--myeloid crosstalk, with CD4$^+$ T, CD8$^+$ T, and NK cells emerging as broadly connected cell types and B cells and monocytes participating in more selective interactions; reciprocal and unidirectional selections are distinguished graphically.
At the level of directed cell-type interfaces, the selected edges have several well-established biological precedents. CD4$^+$ T cells can shape CD8$^+$ T-cell activation and memory formation, and CD4$^+$ T--B-cell interactions are central to B-cell activation and humoral responses \citep*{laidlaw2016multifaceted,crotty2011follicular,crotty2015brief}. The reciprocal CD8$^+$ T-to-CD4$^+$ T selection is also biologically plausible, as specialized KIR$^+$CD8$^+$ regulatory T cells can selectively induce apoptosis of pathogenic CD4$^+$ T cells \citep{li2022kir+}. B cells, acting as antigen-presenting cells, can modulate CD8$^+$ T-cell responses \citep{rastogi2022role,vanmeerhaeghe2023regulation}. In the reverse direction, alloprimed CD8$^+$ T cells have been shown to directly induce apoptosis of alloprimed IgG1$^+$ B cells in a target-cell MHC class-I-dependent manner \citep{zimmerer2014alloprimed}.

The reciprocal selections involving NK cells and T cells are compatible with evidence that NK cells can regulate adaptive CD4$^+$ and CD8$^+$ T-cell responses through cytokine-mediated, antigen-presenting-cell-mediated, and cytotoxic mechanisms \citep{crouse2015nk,pallmer2016recognition}. Conversely, antigen-primed CD4$^+$ T cells can promote NK-cell activation through IL-2- and CD40L-dependent pathways \citep{bihl2010primed}, while reactivation of tissue-resident memory CD8$^+$ T cells can induce local NK-cell effector activation \citep{schenkel2014resident}. The reciprocal NK--B-cell selections likewise have precedents in both directions. NK cells can regulate B-cell activation, differentiation, and antibody responses \citep*{gyurova2020natural}, whereas activated B cells can directly alter NK-cell cytokine expression through CD48--CD244-mediated cell--cell interactions \citep{gao2006bcell}.

The monocyte-related selections are also consistent with established direction-specific interactions. Monocytes and monocyte-derived cells can present antigen to and promote CD4$^+$ T-cell activation \citep*{randolph2008antigen}. Conversely, human CD4$^+$CD25$^+$ regulatory T cells can directly suppress monocyte inflammatory function, whereas activated CD8$^+$ T cells can induce monocyte differentiation toward dendritic cells \citep{taams2005modulation,wirths2002activated}. Finally, the reciprocal NK--monocyte selections are consistent with cytokine- and contact-mediated crosstalk through which monocytes promote NK-cell activation and NK-cell-derived signals, in turn, regulate monocyte activation and differentiation \citep*{knorr2014interplay}.

In summary, these results identify distributional sender--target associations that are consistent with known immune biology.

\section{Discussion}
\label{sec:discussion}

We have presented a Bayesian method for distribution-to-distribution regression with explicit variable selection, motivated by the challenge of inferring cell-cell communication networks from population-scale single-cell data. The key insight of \spider{} is that representing distributions as finite mixtures with shared component parameters reduces the regression problem to a multivariate linear model on subject-specific mixing weights, thereby unlocking classical Bayesian tools for inference and selection.

Several limitations suggest directions for future research. First, the shared-parameter assumption requires that all subjects share a common set of mixture component locations and covariances, with heterogeneity captured solely through mixing weights. While this provides computational tractability and interpretability, it may be restrictive when subpopulation structure varies substantially across subjects. Second, the linear latent regression may not capture complex nonlinear dependencies; nonparametric extensions are a natural direction for future work.

Although our application focused on cell-cell communication, the framework applies to any distributional regression problem in which identifying a sparse set of relevant distributional predictors is of scientific interest.

\section*{Acknowledgments}
During the preparation of this manuscript, the authors utilized AI-assisted technologies. Anthropic's Claude Opus 4.6 was employed for language polishing and editing. OpenAI Codex, using GPT-5.4 and GPT-5.5, was used to assist in debugging and correcting the code. All AI-assisted outputs were reviewed, edited, and verified by the authors, who take full responsibility for the accuracy and integrity of the work.

\bibliographystyle{chicago}
\bibliography{references}

\clearpage
\appendix
\renewcommand{\thesection}{\Alph{section}}
\renewcommand{\thesubsection}{\thesection.\arabic{subsection}}
\renewcommand{\theequation}{\thesection.\arabic{equation}}
\renewcommand{\thetable}{S\arabic{table}}
\renewcommand{\thefigure}{S\arabic{figure}}
\renewcommand{\theHsection}{appendix.\Alph{section}}
\renewcommand{\theHequation}{appendix.\thesection.\arabic{equation}}
\renewcommand{\theHtable}{appendix.S\arabic{table}}
\renewcommand{\theHfigure}{appendix.S\arabic{figure}}
\setcounter{table}{0}
\setcounter{figure}{0}
\numberwithin{equation}{section}

\section{Proof of Proposition~\ref{prop:mixture_identifiability}}
\label{supp:proof_identifiability}

\begin{proof}[Proof of Proposition~\ref{prop:mixture_identifiability}]
For each subject \(i\), identifiability of finite mixtures of
multivariate Gaussian distributions
\citep{yakowitz1968identifiability} implies that there exists a
permutation \(\sigma_i\in S_L\) such that
\[
(\widetilde{\bm{\mu}}_{\sigma_i(\ell)},
 \widetilde{\bm{\Sigma}}_{\sigma_i(\ell)})
=
(\bm{\mu}_\ell,\bm{\Sigma}_\ell),
\qquad
\widetilde{\pi}_{i,\sigma_i(\ell)}
=
\pi_{i\ell},
\qquad \ell=1,\ldots,L.
\]
Let \(\sigma=\sigma_1\). Because the component dictionaries are shared
across subjects and the component parameter pairs are pairwise
distinct, the permutation matching the two dictionaries is unique.
Consequently, \(\sigma_i=\sigma\) for every \(i\), which establishes
the result.
\end{proof}

\section{MCMC Inference Details}
\label{supp:mcmc_inference}
\label{app:mcmc}

In this section, we provide the detailed conditional update formulas and implementation details.

\subsection{GMM Component Updates}
\label{app:gmm_updates}

This section summarizes the conditional updates for the Gaussian mixture kernels used to represent the response and predictor distributions.

\paragraph{Generic Normal--Inverse-Wishart update.}
Consider observations assigned to a single Gaussian component,
\[
    \mathcal D=\{\mathbf y_1,\ldots,\mathbf y_n\}, \qquad \mathbf y_r\in\mathbb R^d,
\]
with
\[
    \mathbf y_r\mid\bm\mu,\bm\Sigma \stackrel{\mathrm{ind}}{\sim} \mathcal N_d(\bm\mu,\bm\Sigma), \qquad r=1,\ldots,n.
\]
The conjugate prior is
\[
    \bm\Sigma\sim\mathcal{IW}_d(\nu_0,\bm\Psi_0), \qquad \bm\mu\mid\bm\Sigma \sim \mathcal N_d\left( \bm\xi_0, \frac{1}{\kappa_0}\bm\Sigma \right), \qquad \kappa_0>0.
\]
For $n>0$, let
\[
    \bar{\mathbf y} = \frac{1}{n}\sum_{r=1}^n\mathbf y_r, \qquad \mathbf S = \sum_{r=1}^n (\mathbf y_r-\bar{\mathbf y}) (\mathbf y_r-\bar{\mathbf y})^\top.
\]
The posterior hyperparameters are
\begin{align}
    \kappa_n &= \kappa_0+n, & \nu_n &= \nu_0+n, \nonumber\\
    \bm\mu_n &= \frac{\kappa_0\bm\xi_0+n\bar{\mathbf y}} {\kappa_n}, & \bm\Psi_n &= \bm\Psi_0 + \mathbf S + \frac{\kappa_0 n}{\kappa_n} (\bar{\mathbf y}-\bm\xi_0) (\bar{\mathbf y}-\bm\xi_0)^\top.
    \label{eq:supp_gmm_niw_hyper}
\end{align}
Thus
\begin{align}
    \bm\Sigma\mid\mathcal D &\sim \mathcal{IW}_d(\nu_n,\bm\Psi_n), \label{eq:supp_gmm_sigma_post}\\
    \bm\mu\mid\bm\Sigma,\mathcal D &\sim \mathcal N_d\left( \bm\mu_n, \frac{1}{\kappa_n}\bm\Sigma \right).
    \label{eq:supp_gmm_mu_post}
\end{align}

\paragraph{Response mixture component update.}
Let $z^Y_{i,c}\in\{1,\ldots,K\}$, \(c=1,\ldots,N_i^Y\), denote the response-component allocation for observation \(c\) in subject \(i\). For component $k$, define
\[
    \mathcal I^Y_k = \{(i,c):z^Y_{i,c}=k\}, \qquad n^Y_k=|\mathcal I^Y_k|, \qquad \mathcal D^Y_k = \{\mathbf y_{i,c}:(i,c)\in\mathcal I^Y_k\}.
\]
For \(n_k^Y>0\), define
\[
    \bar{\mathbf y}_k = \frac{1}{n_k^Y} \sum_{(i,c)\in\mathcal I_k^Y}\mathbf y_{i,c}, \qquad \mathbf S_k^Y = \sum_{(i,c)\in\mathcal I_k^Y} (\mathbf y_{i,c}-\bar{\mathbf y}_k) (\mathbf y_{i,c}-\bar{\mathbf y}_k)^\top.
\]
Applying \eqref{eq:supp_gmm_niw_hyper} with \(d=d_Y\), \(n=n_k^Y\), and prior hyperparameters \((\bm\xi_Y,\kappa_Y,\nu_Y,\bm\Psi_Y)\) gives
\begin{align}
    \kappa^Y_{k,n} &= \kappa_Y+n_k^Y, & \nu^Y_{k,n} &= \nu_Y+n_k^Y, \nonumber\\
    \bm\mu^Y_{k,n} &= \frac{\kappa_Y\bm\xi_Y+n_k^Y\bar{\mathbf y}_k} {\kappa^Y_{k,n}}, & \bm\Psi^Y_{k,n} &= \bm\Psi_Y + \mathbf S_k^Y + \frac{\kappa_Y n_k^Y}{\kappa^Y_{k,n}} (\bar{\mathbf y}_k-\bm\xi_Y) (\bar{\mathbf y}_k-\bm\xi_Y)^\top.
    \label{eq:supp_response_niw_hyper}
\end{align}
Thus
\[
    \bm\Sigma^Y_k \mid \mathcal D_k^Y \sim \mathcal{IW}_{d_Y} (\nu^Y_{k,n},\bm\Psi^Y_{k,n}),
    \qquad
    \bm\mu^Y_k \mid \bm\Sigma^Y_k,\mathcal D_k^Y \sim \mathcal N_{d_Y} \left( \bm\mu^Y_{k,n}, \frac{1}{\kappa^Y_{k,n}}\bm\Sigma^Y_k \right).
\]
After the response component parameters are sampled, the response components and their allocations are relabeled according to the first coordinate of $\bm\mu^Y_k$ to maintain a stable label convention.

\paragraph{Predictor mixture component update.}
For predictor block $g$, let $z^{(g)}_{i,c}\in\{1,\ldots,J_g\}$, $c=1,\ldots,N_i^{(g)}$, denote the allocation of observation $c$ in subject $i$. For component $j$, define
\[
    \mathcal I^{(g)}_j = \{(i,c):z^{(g)}_{i,c}=j\}, \qquad n^{(g)}_j=|\mathcal I^{(g)}_j|, \qquad \mathcal D^{(g)}_j = \{\mathbf x^{(g)}_{i,c}:(i,c)\in\mathcal I^{(g)}_j\}.
\]
For \(n_j^{(g)}>0\), define
\[
    \bar{\mathbf x}^{(g)}_j = \frac{1}{n_j^{(g)}} \sum_{(i,c)\in\mathcal I_j^{(g)}} \mathbf x^{(g)}_{i,c},
    \qquad
    \mathbf S_j^{(g)} = \sum_{(i,c)\in\mathcal I_j^{(g)}} \left( \mathbf x^{(g)}_{i,c} - \bar{\mathbf x}^{(g)}_j \right) \left( \mathbf x^{(g)}_{i,c} - \bar{\mathbf x}^{(g)}_j \right)^\top.
\]
Ignoring the ordering constraint, the Normal--Inverse-Wishart update with prior hyperparameters \((\bm\xi_g,\kappa_g,\nu_g,\bm\Psi_g)\) gives
\begin{align}
    \kappa^{(g)}_{j,n} &= \kappa_g+n_j^{(g)}, & \nu^{(g)}_{j,n} &= \nu_g+n_j^{(g)}, \nonumber\\
    \widetilde{\bm\mu}^{(g)}_{j,n} &= \frac{ \kappa_g\bm\xi_g + n_j^{(g)}\bar{\mathbf x}^{(g)}_j } {\kappa^{(g)}_{j,n}}, & \widetilde{\bm\Psi}^{(g)}_{j,n} &= \bm\Psi_g + \mathbf S_j^{(g)} + \frac{\kappa_g n_j^{(g)}}{\kappa^{(g)}_{j,n}} \left( \bar{\mathbf x}^{(g)}_j-\bm\xi_g \right) \left( \bar{\mathbf x}^{(g)}_j-\bm\xi_g \right)^\top.
    \label{eq:supp_predictor_niw_hyper}
\end{align}
Thus
\[
    \bm\Sigma^X_{g,j} \mid \mathcal D_j^{(g)} \sim \mathcal{IW}_{d_g} \left( \nu^{(g)}_{j,n}, \widetilde{\bm\Psi}^{(g)}_{j,n} \right).
\]
Conditional on \(\bm\Sigma^X_{g,j}\), the unconstrained posterior for the mean is
\[
    \bm\mu^X_{g,j} \mid \bm\Sigma^X_{g,j},\mathcal D_j^{(g)} \sim \mathcal N_{d_g} \left( \widetilde{\bm\mu}^{(g)}_{j,n}, \mathbf A^X_{g,j,n} \right), \qquad \mathbf A^X_{g,j,n} = \frac{1}{\kappa^{(g)}_{j,n}} \bm\Sigma^X_{g,j}.
\]
To identify the predictor mixture components, we impose the ordering constraint
\begin{equation}
\label{eq:supp_gmm_order_constraint}
    \mu^X_{g,1,1} < \mu^X_{g,2,1} < \cdots < \mu^X_{g,J_g,1},
\end{equation}
where the last subscript denotes the first coordinate of the component mean. The constraint is imposed within the Gibbs update of each component mean, conditional on the current neighboring component means.

For component \(j\), partition the unconstrained posterior mean and covariance as
\[
    \widetilde{\bm\mu}^{(g)}_{j,n} =
    \begin{pmatrix}
        m_1\\
        \mathbf m_{-1}
    \end{pmatrix},
    \qquad
    \mathbf A^X_{g,j,n} =
    \begin{pmatrix}
        a_{11} & \mathbf a_{1,-1}\\
        \mathbf a_{-1,1} & \mathbf A^X_{-1,-1}
    \end{pmatrix}.
\]
The allowable interval for the first coordinate is
\[
    \ell_{g,j} =
    \begin{cases}
        -\infty, & j=1,\\
        \mu^X_{g,j-1,1}, & j>1,
    \end{cases}
    \qquad
    u_{g,j} =
    \begin{cases}
        \mu^X_{g,j+1,1}, & j<J_g,\\
        +\infty, & j=J_g.
    \end{cases}
\]
Thus
\[
    \mu^X_{g,j,1} \mid \bm\Sigma^X_{g,j}, \mathcal D_j^{(g)}, \bm\mu^X_{g,-j} \sim \mathcal{TN}_{(\ell_{g,j},u_{g,j})}(m_1,a_{11}),
\]
where \(\bm\mu^X_{g,-j}\) denotes the current component means in block \(g\) except for component \(j\), and \(\mathcal{TN}_{(a,b)}(m,s^2)\) denotes a normal distribution with mean \(m\) and variance \(s^2\), truncated to \((a,b)\).

Given the sampled first coordinate \(\mu^{\star}_{g,j,1}\), define
\begin{align*}
    \mathbf m^{\star}_{g,j,-1}
    &= \mathbf m_{-1}
       + \mathbf a_{-1,1}a_{11}^{-1}
         \bigl(\mu^{\star}_{g,j,1}-m_1\bigr), \\
    \mathbf A^{\star}_{g,j,-1}
    &= \mathbf A^X_{-1,-1}
       - \mathbf a_{-1,1}a_{11}^{-1}
         \mathbf a_{1,-1}.
\end{align*}
The remaining coordinates are sampled according to
\[
    \bm\mu^X_{g,j,-1}
     \mid \mu^X_{g,j,1}=\mu^{\star}_{g,j,1},
           \bm\Sigma^X_{g,j},\mathcal D_j^{(g)}
     \sim
     \mathcal N_{d_g-1}
     \left(
       \mathbf m^{\star}_{g,j,-1},
       \mathbf A^{\star}_{g,j,-1}
     \right).
\]

\paragraph{Allocation updates.}
The component allocations are updated from their categorical full conditional distributions. For a response observation \(\mathbf y_{i,c}\), define the unnormalized weights
\[
    r^Y_{i,c,k} = \pi^Y_{i,k} \mathcal N_{d_Y} \left( \mathbf y_{i,c} \mid \bm\mu^Y_k, \bm\Sigma^Y_k \right), \qquad k=1,\ldots,K.
\]
Then
\begin{equation}
\label{eq:supp_response_alloc_update}
    \Pr(z^Y_{i,c}=k\mid \mathbf y_{i,c},\bm\pi_i^Y, \{\bm\mu^Y_\ell,\bm\Sigma^Y_\ell\}_{\ell=1}^{K}) = \frac{r^Y_{i,c,k}} {\sum_{\ell=1}^{K} r^Y_{i,c,\ell}}.
\end{equation}

Similarly, for predictor block \(g\) and predictor observation \(\mathbf x^{(g)}_{i,c}\), define
\[
    r^{(g)}_{i,c,j} = \pi^{(g)}_{i,j} \mathcal N_{d_g} \left( \mathbf x^{(g)}_{i,c} \mid \bm\mu^X_{g,j}, \bm\Sigma^X_{g,j} \right), \qquad j=1,\ldots,J_g.
\]
The predictor allocation is sampled according to
\begin{equation}
\label{eq:supp_predictor_alloc_update}
    \Pr(z^{(g)}_{i,c}=j\mid \mathbf x^{(g)}_{i,c},\bm\pi_i^{(g)}, \{\bm\mu^X_{g,\ell},\bm\Sigma^X_{g,\ell}\}_{\ell=1}^{J_g}) = \frac{r^{(g)}_{i,c,j}} {\sum_{\ell=1}^{J_g} r^{(g)}_{i,c,\ell}}.
\end{equation}
The subject-specific weights are deterministic functions of the current latent coordinates:
\[
    \bm\pi_i^Y = \operatorname{ALR}^{-1}_K(\mathbf U_i), \qquad \bm\pi_i^{(g)} = \operatorname{SB}^{-1}_{J_g}(\mathbf V_i^{(g)}).
\]

\paragraph{Gaussian hyperprior update for predictor coordinates.}
For each predictor block \(g\) with \(J_g>1\), let
\[
    d_{V,g}=J_g-1, \qquad \mathcal V_g = \{\mathbf V_i^{(g)}:i=1,\ldots,N\}.
\]
The subject-specific stick-breaking coordinates satisfy
\[
    \mathbf V_i^{(g)} \mid \bm\mu_V^{(g)},\bm\Sigma_V^{(g)} \stackrel{\mathrm{iid}}{\sim} \mathcal N_{d_{V,g}} \left( \bm\mu_V^{(g)}, \bm\Sigma_V^{(g)} \right).
\]
The hyperprior is
\[
    \bm\Sigma_V^{(g)} \sim \mathcal{IW}_{d_{V,g}} \left( \nu_{V0}^{(g)}, \bm\Psi_{V0}^{(g)} \right), \qquad \bm\mu_V^{(g)} \mid \bm\Sigma_V^{(g)} \sim \mathcal N_{d_{V,g}} \left( \mathbf m_{V0}^{(g)}, \frac{1}{\lambda_{V0}^{(g)}} \bm\Sigma_V^{(g)} \right).
\]
Define the empirical mean and scatter matrix
\[
    \bar{\mathbf V}^{(g)} = \frac{1}{N} \sum_{i=1}^N \mathbf V_i^{(g)}, \qquad \mathbf S_V^{(g)} = \sum_{i=1}^N \left( \mathbf V_i^{(g)} - \bar{\mathbf V}^{(g)} \right) \left( \mathbf V_i^{(g)} - \bar{\mathbf V}^{(g)} \right)^\top.
\]
The Normal--Inverse-Wishart update gives
\begin{align}
    \lambda_{V,n}^{(g)}
    &= \lambda_{V0}^{(g)}+N,
    \qquad
    \nu_{V,n}^{(g)}
    = \nu_{V0}^{(g)}+N,
    \nonumber\\
    \mathbf m_{V,n}^{(g)}
    &=
    \frac{
      \lambda_{V0}^{(g)}\mathbf m_{V0}^{(g)}
      +N\bar{\mathbf V}^{(g)}
    }{
      \lambda_{V,n}^{(g)}
    },
    \nonumber\\
    \bm\Psi_{V,n}^{(g)}
    &=
    \bm\Psi_{V0}^{(g)}+\mathbf S_V^{(g)}
    +\frac{\lambda_{V0}^{(g)}N}
           {\lambda_{V,n}^{(g)}}
     \left(
       \bar{\mathbf V}^{(g)}-\mathbf m_{V0}^{(g)}
     \right)
     \left(
       \bar{\mathbf V}^{(g)}-\mathbf m_{V0}^{(g)}
     \right)^\top.
\end{align}
Thus
\[
    \bm\Sigma_V^{(g)} \mid \mathcal V_g \sim \mathcal{IW}_{d_{V,g}} \left( \nu_{V,n}^{(g)}, \bm\Psi_{V,n}^{(g)} \right),
    \qquad
    \bm\mu_V^{(g)} \mid \bm\Sigma_V^{(g)},\mathcal V_g \sim \mathcal N_{d_{V,g}} \left( \mathbf m_{V,n}^{(g)}, \frac{1}{\lambda_{V,n}^{(g)}} \bm\Sigma_V^{(g)} \right).
\]

\subsection{Predictor-Coordinate Updates}
\label{app:polya_gamma}

The Gaussian mixture updates as described above condition on the current subject-specific mixture weights. We now describe the update of the predictor latent coordinates.

\paragraph{Stick-breaking likelihood.}
For predictor \(g\), let
\[
    \mathbf V_i^{(g)} = \left( V_{i1}^{(g)},\ldots,V_{i,J_g-1}^{(g)} \right)^\top
\]
denote the stick-breaking coordinates for subject \(i\). The corresponding mixture weights are
\[
    \pi_{i,j}^{(g)} = \sigma\!\left(V_{ij}^{(g)}\right) \prod_{\ell<j} \left\{ 1-\sigma\!\left(V_{i\ell}^{(g)}\right) \right\}, \qquad j=1,\ldots,J_g-1,
\]
and
\[
    \pi_{i,J_g}^{(g)} = \prod_{\ell=1}^{J_g-1} \left\{ 1-\sigma\!\left(V_{i\ell}^{(g)}\right) \right\}, \qquad \sigma(x)=\{1+\exp(-x)\}^{-1}.
\]

Let
\[
    \mathbf z_i^{(g)} = \left( z_{i,1}^{(g)},\ldots,z_{i,N_i^{(g)}}^{(g)} \right)
\]
be the allocation vector for the sample of size \(N_i^{(g)}\) from subject \(i\) in predictor block \(g\), where \(z_{i,c}^{(g)}\in\{1,\ldots,J_g\}\). Define the component counts
\[
    n_{ij}^{(g)} = \sum_{c=1}^{N_i^{(g)}} \mathbf 1\{z_{i,c}^{(g)}=j\}, \qquad j=1,\ldots,J_g,
\]
and collect them as
\[
    \mathbf n_i^{(g)} = \left( n_{i1}^{(g)},\ldots,n_{iJ_g}^{(g)} \right)^\top.
\]
The part of the allocation likelihood that depends on \(\mathbf V_i^{(g)}\) depends on the cell-level allocations only through the component counts. Indeed,
\[
    p(\mathbf z_i^{(g)}\mid \mathbf V_i^{(g)}) = \prod_{c=1}^{N_i^{(g)}} \pi^{(g)}_{i,z^{(g)}_{i,c}} = \prod_{j=1}^{J_g} \left( \pi_{i,j}^{(g)} \right)^{n_{ij}^{(g)}} .
\]
Equivalently, the multinomial count likelihood differs from this expression only by a multinomial coefficient that is constant in \(\mathbf V_i^{(g)}\). Thus, for the update of \(\mathbf V_i^{(g)}\), we may write the likelihood in terms of the count vector \(\mathbf n_i^{(g)}\).

Following the logistic stick-breaking decomposition of the multinomial likelihood in \citet*{linderman2015dependent}, define the remaining-stick counts
\[
    N_{ij}^{(g)} = \sum_{\ell=j}^{J_g} n_{i\ell}^{(g)}, \qquad j=1,\ldots,J_g-1.
\]
Then, the allocation likelihood can be written as
\begin{align}
\label{eq:supp_pg_sb_likelihood}
    p(\mathbf n_i^{(g)}\mid \mathbf V_i^{(g)}) &\propto \prod_{j=1}^{J_g-1} \left[ \sigma\!\left(V_{ij}^{(g)}\right) \right]^{n_{ij}^{(g)}} \left[ 1-\sigma\!\left(V_{ij}^{(g)}\right) \right]^{N_{ij}^{(g)}-n_{ij}^{(g)}} \nonumber\\
    &= \prod_{j=1}^{J_g-1} \frac{ \exp\{n_{ij}^{(g)}V_{ij}^{(g)}\} }{ \left[ 1+\exp\{V_{ij}^{(g)}\} \right]^{N_{ij}^{(g)}} }.
\end{align}
Thus, for fixed allocations, the stick-breaking representation expresses the multinomial likelihood as a product of binomial-logit kernels indexed by \(j=1,\ldots,J_g-1\).

\paragraph{P\'olya--Gamma augmentation.}
We apply the P\'olya--Gamma identity of \citet*{polson2013bayesian} to each factor in \eqref{eq:supp_pg_sb_likelihood}, following the multinomial stick-breaking construction of \citet*{linderman2015dependent}. Introduce independent auxiliary variables
\[
    \omega_{ij}^{(g)} \mid V_{ij}^{(g)},\mathbf n_i^{(g)} \sim \mathrm{PG} \left( N_{ij}^{(g)}, V_{ij}^{(g)} \right), \qquad j=1,\ldots,J_g-1.
\]
Define
\[
    \kappa_{ij}^{(g)} = n_{ij}^{(g)} - \frac{1}{2}N_{ij}^{(g)}.
\]
Conditional on the augmented variables, the likelihood contribution from block \(g\) is proportional to
\begin{equation}
\label{eq:supp_pg_block_kernel}
    \exp\left\{ \sum_{j=1}^{J_g-1} \kappa_{ij}^{(g)} V_{ij}^{(g)} - \frac{1}{2} \sum_{j=1}^{J_g-1} \omega_{ij}^{(g)} \left(V_{ij}^{(g)}\right)^2 \right\}.
\end{equation}

Let
\[
    \mathcal I_V = \left\{ (g,j): g=1,\ldots,G,\; j=1,\ldots,J_g-1 \right\},
\]
ordered lexicographically by \(g\) and then \(j\). The stacked predictor coordinate is
\[
    \mathbf V_i = \left( V_{ij}^{(g)} : (g,j)\in\mathcal I_V \right)^\top.
\]
Define
\[
    \bm\kappa_i = \left( \kappa_{ij}^{(g)} : (g,j)\in\mathcal I_V \right)^\top, \qquad \bm\Omega_i = \operatorname{diag} \left\{ \omega_{ij}^{(g)} : (g,j)\in\mathcal I_V \right\}.
\]
Combining the blockwise kernels gives
\begin{equation}
\label{eq:supp_pg_stacked_kernel}
    p(\{\mathbf n_i^{(g)}\}_{g=1}^{G},\bm\omega_i \mid \mathbf V_i) \propto \exp\left\{ \bm\kappa_i^\top\mathbf V_i - \frac{1}{2} \mathbf V_i^\top \bm\Omega_i \mathbf V_i \right\}.
\end{equation}
Therefore, conditional on the P\'olya--Gamma variables, the predictor allocation likelihood is Gaussian in \(\mathbf V_i\).

\paragraph{Gaussian full conditional for predictor coordinates.}
We now combine the P\'olya--Gamma augmented likelihood in \eqref{eq:supp_pg_stacked_kernel} with the Gaussian prior for the predictor coordinates and the Gaussian regression layer. Let
\[
\bm\mu_V = \left( (\bm\mu_V^{(1)})^\top,\ldots,(\bm\mu_V^{(G)})^\top \right)^\top, \qquad \bm\Sigma_V = \operatorname{blkdiag} \left( \bm\Sigma_V^{(1)},\ldots,\bm\Sigma_V^{(G)} \right),
\]
and let \(Q=\sum_{g=1}^G(J_g-1)\). The hierarchical prior gives
\[
    \mathbf V_i \mid \bm\mu_V,\bm\Sigma_V \sim \mathcal N_Q(\bm\mu_V,\bm\Sigma_V),
\]
whose log-kernel in \(\mathbf V_i\) is
\begin{equation}
\label{eq:supp_V_prior_kernel}
    \log p(\mathbf V_i\mid\bm\mu_V,\bm\Sigma_V) = - \frac{1}{2} \mathbf V_i^\top \bm\Sigma_V^{-1} \mathbf V_i + \mathbf V_i^\top \bm\Sigma_V^{-1} \bm\mu_V + C_1,
\end{equation}
where \(C_1\) does not depend on \(\mathbf V_i\).

From \eqref{eq:supp_pg_stacked_kernel}, the augmented predictor-allocation likelihood contributes
\begin{equation}
\label{eq:supp_V_pg_kernel}
    \log p(\{\mathbf n_i^{(g)}\}_{g=1}^G,\bm\omega_i \mid \mathbf V_i) = \bm\kappa_i^\top\mathbf V_i - \frac{1}{2} \mathbf V_i^\top \bm\Omega_i \mathbf V_i + C_2,
\end{equation}
where \(C_2\) is constant in \(\mathbf V_i\).

The response-regression layer is
\[
    \mathbf U_i = \boldsymbol\alpha + \mathbf B^\top\mathbf V_i + \boldsymbol\varepsilon_i, \qquad \boldsymbol\varepsilon_i \sim \mathcal N_q(\mathbf 0,\bm\Sigma_U),
\]
where \(\mathbf B\) is the full \(Q\times q\) matrix formed by padding \(\mathbf B_{(\bm\gamma)}\) with zero rows for inactive predictor groups. The intercept \(\boldsymbol\alpha\) is sampled in the uncollapsed regression step and is conditioned upon in the present update. Thus, with
\[
    \mathbf r_i=\mathbf U_i-\boldsymbol\alpha,
\]
the part of the regression log-likelihood involving \(\mathbf V_i\) is
\begin{align}
\label{eq:supp_V_regression_kernel}
    \log p(\mathbf U_i\mid \mathbf V_i,\boldsymbol\alpha,\mathbf B,\bm\Sigma_U) &= - \frac{1}{2} \left( \mathbf r_i-\mathbf B^\top\mathbf V_i \right)^\top \bm\Sigma_U^{-1} \left( \mathbf r_i-\mathbf B^\top\mathbf V_i \right) + C_3 \nonumber\\
    &= - \frac{1}{2} \mathbf V_i^\top \mathbf B\bm\Sigma_U^{-1}\mathbf B^\top \mathbf V_i + \mathbf V_i^\top \mathbf B\bm\Sigma_U^{-1} (\mathbf U_i-\boldsymbol\alpha) + C_4.
\end{align}
The second equality follows by expanding the quadratic form and retaining only the terms that depend on \(\mathbf V_i\).

Multiplying the three kernels in \eqref{eq:supp_V_prior_kernel}--\eqref{eq:supp_V_regression_kernel} and collecting the terms that depend on \(\mathbf V_i\) gives
\[
    \log p(\mathbf V_i\mid\mathcal C_i) = - \frac{1}{2} \mathbf V_i^\top \bm\Lambda_i \mathbf V_i + \mathbf V_i^\top \bm\eta_i + C,
\]
where \(\mathcal C_i\) denotes the current values of all remaining parameters, allocations, and P\'olya--Gamma variables, and
\begin{align}
\label{eq:supp_V_precision_alpha}
    \bm\Lambda_i &= \bm\Sigma_V^{-1} + \bm\Omega_i + \mathbf B\bm\Sigma_U^{-1}\mathbf B^\top, \\
\label{eq:supp_V_linear_alpha}
    \bm\eta_i &= \bm\Sigma_V^{-1}\bm\mu_V + \bm\kappa_i + \mathbf B\bm\Sigma_U^{-1} (\mathbf U_i-\boldsymbol\alpha).
\end{align}
Since \(\bm\Sigma_V^{-1}\) is positive definite and the remaining two precision contributions are positive semidefinite, \(\bm\Lambda_i\) is positive definite. Completing the square,
\[
    - \frac{1}{2} \mathbf V_i^\top\bm\Lambda_i\mathbf V_i + \mathbf V_i^\top\bm\eta_i = - \frac{1}{2} \left( \mathbf V_i-\bm\Lambda_i^{-1}\bm\eta_i \right)^\top \bm\Lambda_i \left( \mathbf V_i-\bm\Lambda_i^{-1}\bm\eta_i \right) + C',
\]
where \(C'\) is constant in \(\mathbf V_i\). Therefore
\begin{equation}
\label{eq:supp_V_full_conditional_alpha}
    \mathbf V_i \mid \mathcal C_i \sim \mathcal N_Q \left( \bm\Lambda_i^{-1}\bm\eta_i,\, \bm\Lambda_i^{-1} \right).
\end{equation}
When no predictor group is active, \(\mathbf B=\mathbf 0\), and the regression feedback term in \eqref{eq:supp_V_precision_alpha}--\eqref{eq:supp_V_linear_alpha} drops out.

\subsection{Collapsed--Uncollapsed Sampling for the Regression Layer}
\label{app:cu_sampling}

In this section, we derive the posterior conditionals underlying the collapsed--uncollapsed update of the regression layer.

\paragraph{Intercept marginalization.}
Let \(q=K-1\) be the dimension of the response coordinate. Conditional on the inclusion vector \(\bm\gamma\), let \(\mathbf V_{(\bm\gamma)}\) denote the \(N\times p_\gamma\) matrix of active predictor coordinates, and let \(\mathbf B_{(\bm\gamma)}\) be the corresponding \(p_\gamma\times q\) coefficient matrix. The regression layer is
\begin{equation}
\label{eq:supp_regression_model}
    \mathbf U = \mathbf 1_N\boldsymbol\alpha^\top + \mathbf V_{(\bm\gamma)} \mathbf B_{(\bm\gamma)} + \mathbf E, \qquad \mathbf E\mid\bm\Sigma_U \sim \mathcal{MN}_{N\times q} (\mathbf 0,\mathbf I_N,\bm\Sigma_U).
\end{equation}
The active regression rows have the conjugate prior
\[
    \mathbf B_{(\bm\gamma)} \mid \bm\Sigma_U,\bm\gamma \sim \mathcal{MN}_{p_\gamma\times q} \left( \mathbf 0, \mathbf H_{(\bm\gamma)}, \bm\Sigma_U \right),
    \qquad
    \bm\Sigma_U \sim \mathcal{IW}_q(\nu_U,\bm\Psi_U).
\]
Rows corresponding to inactive predictor groups are fixed at zero under the selection prior.

For the intercept, we use the conjugate prior
\[
    \boldsymbol\alpha \mid \bm\Sigma_U \sim \mathcal N_q \left( \boldsymbol\alpha_0, h_\alpha\bm\Sigma_U \right),
\]
and take \(h_\alpha\to\infty\). Define
\[
    \mathbf C_N = \mathbf I_N - \frac{1}{N}\mathbf 1_N\mathbf 1_N^\top, \qquad \widetilde{\mathbf U} = \mathbf C_N\mathbf U, \qquad \widetilde{\mathbf V}_{(\bm\gamma)} = \mathbf C_N\mathbf V_{(\bm\gamma)},
\]
and let
\[
    \bar{\mathbf U} = \frac{1}{N}\mathbf U^\top\mathbf 1_N, \qquad \bar{\mathbf V}_{(\bm\gamma)} = \frac{1}{N} \mathbf V_{(\bm\gamma)}^\top\mathbf 1_N.
\]
For fixed \(\mathbf B_{(\bm\gamma)}\), set
\[
    \mathbf d_\gamma = \bar{\mathbf U} - \mathbf B_{(\bm\gamma)}^\top \bar{\mathbf V}_{(\bm\gamma)}.
\]
Then the residual matrix in \eqref{eq:supp_regression_model} decomposes as
\[
    \mathbf U - \mathbf 1_N\boldsymbol\alpha^\top - \mathbf V_{(\bm\gamma)}\mathbf B_{(\bm\gamma)} = \left( \widetilde{\mathbf U} - \widetilde{\mathbf V}_{(\bm\gamma)} \mathbf B_{(\bm\gamma)} \right) + \mathbf 1_N (\mathbf d_\gamma-\boldsymbol\alpha)^\top.
\]
The two terms on the right-hand side are orthogonal in row space, because both \(\widetilde{\mathbf U}\) and \(\widetilde{\mathbf V}_{(\bm\gamma)}\) have column means zero. Therefore
\begin{align}
\label{eq:supp_intercept_decomposition}
& \left( \mathbf U - \mathbf 1_N\boldsymbol\alpha^\top - \mathbf V_{(\bm\gamma)}\mathbf B_{(\bm\gamma)} \right)^\top \left( \mathbf U - \mathbf 1_N\boldsymbol\alpha^\top - \mathbf V_{(\bm\gamma)}\mathbf B_{(\bm\gamma)} \right) \nonumber\\
&\quad = \left( \widetilde{\mathbf U} - \widetilde{\mathbf V}_{(\bm\gamma)} \mathbf B_{(\bm\gamma)} \right)^\top \left( \widetilde{\mathbf U} - \widetilde{\mathbf V}_{(\bm\gamma)} \mathbf B_{(\bm\gamma)} \right) + N (\boldsymbol\alpha-\mathbf d_\gamma) (\boldsymbol\alpha-\mathbf d_\gamma)^\top.
\end{align}

Combining the Gaussian likelihood with the intercept prior, the terms depending on \(\boldsymbol\alpha\) are
\[
    N (\boldsymbol\alpha-\mathbf d_\gamma) (\boldsymbol\alpha-\mathbf d_\gamma)^\top + h_\alpha^{-1} (\boldsymbol\alpha-\boldsymbol\alpha_0) (\boldsymbol\alpha-\boldsymbol\alpha_0)^\top.
\]
Let
\[
    d_h=N+h_\alpha^{-1}, \qquad \boldsymbol\alpha_h = \frac{ N\mathbf d_\gamma + h_\alpha^{-1}\boldsymbol\alpha_0 }{ d_h }.
\]
Completing the square gives
\begin{align}
\label{eq:supp_alpha_square}
& N (\boldsymbol\alpha-\mathbf d_\gamma) (\boldsymbol\alpha-\mathbf d_\gamma)^\top + h_\alpha^{-1} (\boldsymbol\alpha-\boldsymbol\alpha_0) (\boldsymbol\alpha-\boldsymbol\alpha_0)^\top \nonumber\\
&\quad = d_h (\boldsymbol\alpha-\boldsymbol\alpha_h) (\boldsymbol\alpha-\boldsymbol\alpha_h)^\top + \frac{ N h_\alpha^{-1} }{ N+h_\alpha^{-1} } (\mathbf d_\gamma-\boldsymbol\alpha_0) (\mathbf d_\gamma-\boldsymbol\alpha_0)^\top.
\end{align}
The second term on the right of \eqref{eq:supp_alpha_square} vanishes as \(h_\alpha\to\infty\). Define the centered residual matrix
\[
    \widetilde{\mathbf R}_\gamma
    =
    \widetilde{\mathbf U}
    -
    \widetilde{\mathbf V}_{(\bm\gamma)}
    \mathbf B_{(\bm\gamma)}.
\]
Integrating over \(\boldsymbol\alpha\) in this limit therefore leaves the centered Gaussian likelihood
\begin{equation}
\label{eq:supp_centered_regression_likelihood}
    p\left(
       \mathbf U\mid
       \mathbf V,\mathbf B_{(\bm\gamma)},
       \bm\Sigma_U,\bm\gamma
     \right)
    \propto
    |\bm\Sigma_U|^{-N/2}
    \exp\!\left\{
       -\frac12
       \operatorname{tr}\!\left(
          \bm\Sigma_U^{-1}
          \widetilde{\mathbf R}_\gamma^\top
          \widetilde{\mathbf R}_\gamma
       \right)
    \right\}.
\end{equation}
Thus the collapsed regression calculations use the centered sufficient statistics \(\widetilde{\mathbf U}\) and \(\widetilde{\mathbf V}_{(\bm\gamma)}\), recomputed at each MCMC iteration.

The same calculation gives the uncollapsed conditional distribution of the intercept. In the vague-intercept limit,
\begin{equation}
\label{eq:supp_alpha_conditional}
    \boldsymbol\alpha \mid \mathbf U,\mathbf V,\mathbf B,\bm\Sigma_U,\bm\gamma \sim \mathcal N_q \left( \bar{\mathbf U} - \mathbf B_{(\bm\gamma)}^\top \bar{\mathbf V}_{(\bm\gamma)}, \frac{1}{N}\bm\Sigma_U \right).
\end{equation}
This draw is used in the predictor-coordinate update in Appendix~\ref{app:polya_gamma}.

\paragraph{Marginal likelihood score and collapsed update of $\mathbf{U}$.}
After the intercept has been integrated out as described above, the regression layer is represented by the centered likelihood in \eqref{eq:supp_centered_regression_likelihood}. Thus the remaining collapsed calculations use
\[
    \widetilde{\mathbf U} = \mathbf C_N\mathbf U, \qquad \widetilde{\mathbf V}_{(\bm\gamma)} = \mathbf C_N\mathbf V_{(\bm\gamma)}.
\]
For \(p_\gamma>0\), define
\begin{equation}
\label{eq:supp_K_gamma}
    \mathbf K_\gamma = \widetilde{\mathbf V}_{(\bm\gamma)}^\top \widetilde{\mathbf V}_{(\bm\gamma)} + \mathbf H_{(\bm\gamma)}^{-1},
\end{equation}
and
\begin{equation}
\label{eq:supp_Bhat_gamma}
    \widehat{\mathbf B}_{N,\gamma} = \mathbf K_\gamma^{-1} \widetilde{\mathbf V}_{(\bm\gamma)}^\top \widetilde{\mathbf U}.
\end{equation}
Completing the square in the active coefficient block gives
\begin{align}
\label{eq:supp_centered_B_square}
& \left( \widetilde{\mathbf U} - \widetilde{\mathbf V}_{(\bm\gamma)} \mathbf B_{(\bm\gamma)} \right)^\top \left( \widetilde{\mathbf U} - \widetilde{\mathbf V}_{(\bm\gamma)} \mathbf B_{(\bm\gamma)} \right) + \mathbf B_{(\bm\gamma)}^\top \mathbf H_{(\bm\gamma)}^{-1} \mathbf B_{(\bm\gamma)} \nonumber\\
&\quad = \left( \mathbf B_{(\bm\gamma)} - \widehat{\mathbf B}_{N,\gamma} \right)^\top \mathbf K_\gamma \left( \mathbf B_{(\bm\gamma)} - \widehat{\mathbf B}_{N,\gamma} \right) + \widetilde{\mathbf U}^\top \widetilde{\mathbf U} - \widetilde{\mathbf U}^\top \widetilde{\mathbf V}_{(\bm\gamma)} \mathbf K_\gamma^{-1} \widetilde{\mathbf V}_{(\bm\gamma)}^\top \widetilde{\mathbf U}.
\end{align}
Let
\begin{equation}
\label{eq:supp_S_gamma}
    \mathbf S_\gamma = \widetilde{\mathbf U}^\top \widetilde{\mathbf U} - \widetilde{\mathbf U}^\top \widetilde{\mathbf V}_{(\bm\gamma)} \mathbf K_\gamma^{-1} \widetilde{\mathbf V}_{(\bm\gamma)}^\top \widetilde{\mathbf U}.
\end{equation}
Using \eqref{eq:supp_centered_B_square}, the centered likelihood multiplied by the matrix-normal prior for \(\mathbf B_{(\bm\gamma)}\) has kernel
\begin{align}
\label{eq:supp_likelihood_times_B_prior}
& p_c(\mathbf U\mid \mathbf V,\mathbf B_{(\bm\gamma)},\bm\Sigma_U,\bm\gamma)\, p(\mathbf B_{(\bm\gamma)}\mid\bm\Sigma_U,\bm\gamma) \nonumber\\
&\quad \propto |\mathbf H_{(\bm\gamma)}|^{-q/2} |\bm\Sigma_U|^{-(N+p_\gamma)/2} \nonumber\\
&\qquad \times \exp\left\{ -\frac{1}{2} \operatorname{tr} \left[ \bm\Sigma_U^{-1} \left\{ \left( \mathbf B_{(\bm\gamma)} - \widehat{\mathbf B}_{N,\gamma} \right)^\top \mathbf K_\gamma \left( \mathbf B_{(\bm\gamma)} - \widehat{\mathbf B}_{N,\gamma} \right) + \mathbf S_\gamma \right\} \right] \right\}.
\end{align}
The Gaussian integral over the \(p_\gamma\times q\) coefficient matrix contributes the determinant factor
\[
    |\mathbf K_\gamma|^{-q/2} |\bm\Sigma_U|^{p_\gamma/2},
\]
with the normalizing constants canceling those from the matrix-normal prior. Therefore,
\begin{equation}
\label{eq:supp_after_B_integration}
    p_c(\mathbf U\mid \bm\Sigma_U,\mathbf V,\bm\gamma) \propto |\mathbf H_{(\bm\gamma)}|^{-q/2} |\mathbf K_\gamma|^{-q/2} |\bm\Sigma_U|^{-N/2} \exp\left\{ -\frac{1}{2} \operatorname{tr} \left( \bm\Sigma_U^{-1} \mathbf S_\gamma \right) \right\}.
\end{equation}

Combining \eqref{eq:supp_after_B_integration} with \(\bm\Sigma_U\sim\mathcal{IW}_q(\nu_U,\bm\Psi_U)\) gives the kernel
\[
    |\bm\Sigma_U|^{-(N+\nu_U+q+1)/2} \exp\left\{ -\frac{1}{2} \operatorname{tr} \left( \bm\Sigma_U^{-1} \mathbf Q_\gamma \right) \right\},
\]
where
\begin{equation}
\label{eq:supp_Q_gamma}
    \mathbf Q_\gamma = \bm\Psi_U + \mathbf S_\gamma = \bm\Psi_U + \widetilde{\mathbf U}^\top \widetilde{\mathbf U} - \widetilde{\mathbf U}^\top \widetilde{\mathbf V}_{(\bm\gamma)} \mathbf K_\gamma^{-1} \widetilde{\mathbf V}_{(\bm\gamma)}^\top \widetilde{\mathbf U}.
\end{equation}
Integration over \(\bm\Sigma_U\) yields the factor \(|\mathbf Q_\gamma|^{-(N+\nu_U)/2}\). Thus, up to terms that do not depend on \(\bm\gamma\), the intercept-integrated regression score is
\begin{equation}
\label{eq:supp_gamma_score}
    m(\bm\gamma) = - \frac{q}{2}\log\det \mathbf H_{(\bm\gamma)} - \frac{q}{2}\log\det \mathbf K_\gamma - a_N\log\det \mathbf Q_\gamma, \qquad a_N=\frac{N+\nu_U}{2}.
\end{equation}
When \(p_\gamma=0\), the determinant terms involving \(\mathbf H_{(\bm\gamma)}\) and \(\mathbf K_\gamma\) are omitted and
\[
    \mathbf Q_\emptyset = \bm\Psi_U + \widetilde{\mathbf U}^\top \widetilde{\mathbf U}.
\]
The score \(m(\bm\gamma)\) is used for the elementwise update of the inclusion indicators described below.

We next use the same collapsed regression contribution to update \(\mathbf U\). In this block, \(\mathbf V\), \(\bm\gamma\), and hence \(\widetilde{\mathbf V}_{(\bm\gamma)}\) and \(\mathbf K_\gamma\), are held fixed. For a candidate response-coordinate matrix \(\mathbf U^\star\), let \(\mathbf Q_\gamma[\mathbf U^\star]\) denote the matrix in \eqref{eq:supp_Q_gamma} after replacing \(\widetilde{\mathbf U}\) by \(\mathbf C_N\mathbf U^\star\).

Let \(n_{i,k}^Y\) be the response allocation count for component \(k\) in subject \(i\), and let
\[
    \bm\pi_i^Y(\mathbf U_i^\star) = \operatorname{ALR}^{-1}_K(\mathbf U_i^\star).
\]
Combining the response allocation likelihood with the collapsed regression contribution gives
\begin{equation}
\label{eq:supp_collapsed_U_target}
    p(\mathbf U^\star\mid \mathbf z^Y,\mathbf V,\bm\gamma) \propto \prod_{i=1}^N \prod_{k=1}^K \left\{ \pi_{i,k}^Y(\mathbf U_i^\star) \right\}^{n_{i,k}^Y} \, \left| \mathbf Q_\gamma[\mathbf U^\star] \right|^{-a_N}, \qquad a_N=\frac{N+\nu_U}{2}.
\end{equation}

Following the collapsed Laplace strategy of \citet{silverman2022bayesian}, we update \(\mathbf U\) using a Gaussian approximation to this collapsed target. Let \(\mathbf u=\operatorname{vec}(\mathbf U^\star)\), and define
\[
    \ell_U(\mathbf u) = \sum_{i=1}^N \sum_{k=1}^K n_{i,k}^Y \log \pi_{i,k}^Y(\mathbf U_i^\star) - a_N \log\det \mathbf Q_\gamma[\mathbf U^\star],
\]
where \(\mathbf U^\star=\mathbf U^\star(\mathbf u)\). Let
\[
    \widehat{\mathbf u} = \arg\max_{\mathbf u}\ell_U(\mathbf u), \qquad \mathbf H_U = - \left. \nabla_{\mathbf u}^2 \ell_U(\mathbf u) \right|_{\mathbf u=\widehat{\mathbf u}}.
\]
The collapsed response-coordinate update is based on
\begin{equation}
\label{eq:supp_U_laplace_update}
    \mathbf u \mid \mathbf z^Y,\mathbf V,\bm\gamma \approx \mathcal N \left( \widehat{\mathbf u}, \mathbf H_U^{-1} \right).
\end{equation}
After drawing \(\mathbf U\), the sampler updates the inclusion indicators using the marginal score described next and then recovers the regression parameters from their conditional distributions.

\paragraph{Variable selection via marginal score.}
We now describe the update of the inclusion indicators \(\bm\gamma=(\gamma_1,\ldots,\gamma_G)^\top\). Conditioned on the current latent coordinates \((\mathbf U,\mathbf V)\), the regression parameters \((\boldsymbol\alpha,\mathbf B_{(\bm\gamma)},\bm\Sigma_U)\) are integrated out. The resulting marginal score \(m(\bm\gamma)\) was derived in \eqref{eq:supp_gamma_score}.

Let the prior inclusion probabilities be
\[
    \Pr(\gamma_g=1)=w_g, \qquad \Pr(\gamma_g=0)=1-w_g, \qquad 0<w_g<1.
\]
Then, up to a normalizing constant,
\begin{equation}
\label{eq:supp_gamma_posterior_target}
    p(\bm\gamma\mid \mathbf U,\mathbf V) \propto \exp\{m(\bm\gamma)\} \prod_{g=1}^{G} w_g^{\gamma_g} (1-w_g)^{1-\gamma_g}.
\end{equation}
The sampler updates the elements of \(\bm\gamma\) one at a time. For group \(g\), let \(\bm\gamma^{(1)}\) and \(\bm\gamma^{(0)}\) denote the two candidate inclusion vectors obtained by setting the \(g\)th element to 1 and 0, respectively, while holding all other elements fixed. The log posterior odds are
\begin{equation}
\label{eq:supp_gamma_log_odds}
    \Delta_g = m(\bm\gamma^{(1)}) - m(\bm\gamma^{(0)}) + \log\frac{w_g}{1-w_g}.
\end{equation}
Therefore the elementwise Gibbs update is
\begin{equation}
\label{eq:supp_gamma_update_prob}
    \Pr(\gamma_g=1\mid \bm\gamma_{-g},\mathbf U,\mathbf V) = \frac{\exp(\Delta_g)} {1+\exp(\Delta_g)}.
\end{equation}

Posterior inclusion probabilities are computed from the retained MCMC draws:
\[
    \widehat{\operatorname{PIP}}_g = \frac{1}{T} \sum_{t=1}^{T} \gamma_g^{(t)},
\]
where \(T\) is the number of retained posterior draws.

\paragraph{Conditional regression parameter update.}
After the collapsed update of \(\mathbf U\) and the inclusion vector \(\bm\gamma\), the regression parameters are recovered from their conditional posterior distributions. These draws use the centered sufficient statistics and the matrices \(\mathbf K_\gamma\), \(\widehat{\mathbf B}_{N,\gamma}\), and \(\mathbf Q_\gamma\) defined in \eqref{eq:supp_K_gamma}--\eqref{eq:supp_Q_gamma}.

For \(p_\gamma>0\), define
\[
    \mathbf T_\gamma\left(\mathbf B_{(\bm\gamma)}\right)
    =
    \left(
      \mathbf B_{(\bm\gamma)}
      -\widehat{\mathbf B}_{N,\gamma}
    \right)^\top
    \mathbf K_\gamma
    \left(
      \mathbf B_{(\bm\gamma)}
      -\widehat{\mathbf B}_{N,\gamma}
    \right)
    +\mathbf Q_\gamma .
\]
The completed-square representation in \eqref{eq:supp_centered_B_square} gives the posterior kernel
\[
p\left(
  \mathbf B_{(\bm\gamma)},\bm\Sigma_U
  \mid \mathbf U,\mathbf V,\bm\gamma
 \right)
\propto
 |\bm\Sigma_U|^{-(\nu_U+N+p_\gamma+q+1)/2}
 \exp\!\left\{
   -\frac12
   \operatorname{tr}\!\left[
     \bm\Sigma_U^{-1}
     \mathbf T_\gamma
     \left(\mathbf B_{(\bm\gamma)}\right)
   \right]
 \right\}.
\]
This kernel factors into an inverse-Wishart distribution for \(\bm\Sigma_U\) and a matrix-normal distribution for \(\mathbf B_{(\bm\gamma)}\) conditional on \(\bm\Sigma_U\). Thus
\begin{equation}
\label{eq:supp_SigmaU_conditional}
    \bm\Sigma_U \mid \mathbf U,\mathbf V,\bm\gamma \sim \mathcal{IW}_q \left( \nu_U+N, \mathbf Q_\gamma \right),
\end{equation}
and
\begin{equation}
\label{eq:supp_B_conditional}
    \mathbf B_{(\bm\gamma)} \mid \bm\Sigma_U,\mathbf U,\mathbf V,\bm\gamma \sim \mathcal{MN}_{p_\gamma\times q} \left( \widehat{\mathbf B}_{N,\gamma}, \mathbf K_\gamma^{-1}, \bm\Sigma_U \right).
\end{equation}
The full \(Q\times q\) coefficient matrix used in the predictor-coordinate update is formed by padding \(\mathbf B_{(\bm\gamma)}\) with zero rows for inactive predictor groups.

The intercept is then drawn from the conditional distribution in \eqref{eq:supp_alpha_conditional}.

When \(p_\gamma=0\), the coefficient block is absent and the full coefficient matrix is \(\mathbf B=\mathbf 0\). In this case
\[
    \mathbf Q_\emptyset = \bm\Psi_U + \widetilde{\mathbf U}^\top \widetilde{\mathbf U},
    \qquad
    \bm\Sigma_U \mid \mathbf U,\mathbf V,\bm\gamma \sim \mathcal{IW}_q \left( \nu_U+N, \mathbf Q_\emptyset \right),
    \qquad
    \boldsymbol\alpha \mid \mathbf U,\bm\Sigma_U,\bm\gamma \sim \mathcal N_q \left( \bar{\mathbf U}, \frac{1}{N}\bm\Sigma_U \right).
\]

\subsection{Posterior Prediction}
\label{app:posterior_prediction}

We describe prediction for a single new subject with observed predictor distributions. Let \(t\) index a retained posterior draw from the training sampler. This draw contains the predictor mixture parameters, the response mixture parameters, the selected regression coefficients \(\mathbf B^{(t)}\), the residual covariance \(\bm\Sigma_U^{(t)}\), the inclusion vector \(\bm\gamma^{(t)}\), and the centering quantities \(\bar{\mathbf U}^{(t)}\) and \(\bar{\mathbf V}^{(t)}\).

For predictor block \(g\), let
\[
    \mathcal X_{\mathrm{new}}^{(g)} = \left\{ \mathbf x_{\mathrm{new},1}^{(g)},\ldots, \mathbf x_{\mathrm{new},N_{\mathrm{new}}^{(g)}}^{(g)} \right\}
\]
denote the observed predictor sample for the new subject, where \(N_{\mathrm{new}}^{(g)}\) is the corresponding sample size in block \(g\). Conditional on posterior draw \(t\), write
\[
    \pi_{\mathrm{new},j}^{(g,t)}(\mathbf v) = \left[ \operatorname{SB}_{J_g}^{-1}(\mathbf v) \right]_j, \qquad j=1,\ldots,J_g,
\]
for the stick-breaking mixture weights induced by a candidate coordinate \(\mathbf v\in\mathbb R^{J_g-1}\). The marginal conditional target for the new predictor coordinate is
\begin{equation}
\label{eq:supp_Vnew_target}
\begin{split}
    p_t\!\left( \mathbf V_{\mathrm{new}}^{(g)} = \mathbf v \mid \mathcal X_{\mathrm{new}}^{(g)} \right) &\propto \mathcal N_{J_g-1} \left( \mathbf v \mid \bm\mu_V^{(g,t)}, \bm\Sigma_V^{(g,t)} \right) \\
    &\quad\times \prod_{c=1}^{N_{\mathrm{new}}^{(g)}} \sum_{j=1}^{J_g} \pi_{\mathrm{new},j}^{(g,t)}(\mathbf v) \, \mathcal N_{d_g} \left( \mathbf x_{\mathrm{new},c}^{(g)} \mid \bm\mu_{g,j}^{X(t)}, \bm\Sigma_{g,j}^{X(t)} \right),
\end{split}
\end{equation}
where \(p_t\) denotes conditioning on the retained training draw \(t\). Because the target in \eqref{eq:supp_Vnew_target} is nonconjugate, we use the same stick-breaking P\'olya--Gamma representation as described above.

Introduce latent allocations
\[
    z_{\mathrm{new},c}^{(g,t)} \in \{1,\ldots,J_g\}, \qquad c=1,\ldots,N_{\mathrm{new}}^{(g)}.
\]
For prediction, we use a single P\'olya--Gamma refinement update for \(\mathbf V_{\mathrm{new}}^{(g,t)}\). We initialize
\[
    \mathbf V_{\mathrm{new},0}^{(g,t)} \sim \mathcal N_{J_g-1} \left( \bm\mu_V^{(g,t)}, \bm\Sigma_V^{(g,t)} \right),
    \qquad
    \bm\pi_{\mathrm{new},0}^{(g,t)} = \operatorname{SB}_{J_g}^{-1} \left( \mathbf V_{\mathrm{new},0}^{(g,t)} \right).
\]
Conditional on this initial value, define the unnormalized allocation weights
\[
    r_{\mathrm{new},c,j}^{(g,t)}
    =
      \pi_{\mathrm{new},0,j}^{(g,t)}
      \mathcal N_{d_g}\!\left(
        \mathbf x_{\mathrm{new},c}^{(g)}
        \mid
        \bm\mu_{g,j}^{X(t)},
        \bm\Sigma_{g,j}^{X(t)}
      \right),
    \qquad j=1,\ldots,J_g .
\]
The latent allocation is then sampled according to
\[
\Pr_t\!\left(
 z_{\mathrm{new},c}^{(g,t)}=j
 \mid
 \mathbf V_{\mathrm{new},0}^{(g,t)},
 \mathbf x_{\mathrm{new},c}^{(g)}
\right)
=
\frac{
 r_{\mathrm{new},c,j}^{(g,t)}
}{
 \sum_{\ell=1}^{J_g}
 r_{\mathrm{new},c,\ell}^{(g,t)}
},
\qquad j=1,\ldots,J_g .
\]
Given these allocations, define
\[
    n_{\mathrm{new},j}^{(g,t)} = \sum_{c=1}^{N_{\mathrm{new}}^{(g)}} \mathbf 1 \left\{ z_{\mathrm{new},c}^{(g,t)}=j \right\}, \qquad j=1,\ldots,J_g,
\]
and
\[
    N_{\mathrm{new},j}^{(g,t)} = \sum_{\ell=j}^{J_g} n_{\mathrm{new},\ell}^{(g,t)}, \qquad j=1,\ldots,J_g-1.
\]
For each \(j=1,\ldots,J_g-1\), draw
\[
    \omega_{\mathrm{new},j}^{(g,t)} \mid V_{\mathrm{new},0,j}^{(g,t)}, \{z_{\mathrm{new},c}^{(g,t)}\} \sim \mathrm{PG} \left( N_{\mathrm{new},j}^{(g,t)}, V_{\mathrm{new},0,j}^{(g,t)} \right),
\]
and set
\[
    \kappa_{\mathrm{new},j}^{(g,t)} = n_{\mathrm{new},j}^{(g,t)} - \frac{1}{2} N_{\mathrm{new},j}^{(g,t)}.
\]
Let
\[
    \bm\Omega_{\mathrm{new}}^{(g,t)} = \operatorname{diag} \left( \omega_{\mathrm{new},1}^{(g,t)},\ldots, \omega_{\mathrm{new},J_g-1}^{(g,t)} \right),
    \qquad
    \bm\kappa_{\mathrm{new}}^{(g,t)} = \left( \kappa_{\mathrm{new},1}^{(g,t)},\ldots, \kappa_{\mathrm{new},J_g-1}^{(g,t)} \right)^\top.
\]

Conditional on the sampled allocations and P\'olya--Gamma variables, the augmented stick-breaking likelihood contributes the kernel
\[
    \exp\left\{ \left(\bm\kappa_{\mathrm{new}}^{(g,t)}\right)^\top \mathbf V_{\mathrm{new}}^{(g,t)} - \frac{1}{2} \left(\mathbf V_{\mathrm{new}}^{(g,t)}\right)^\top \bm\Omega_{\mathrm{new}}^{(g,t)} \mathbf V_{\mathrm{new}}^{(g,t)} \right\}.
\]
Combining this kernel with the Gaussian prior under posterior draw \(t\) gives
\[
    \log p\!\left( \mathbf V_{\mathrm{new}}^{(g,t)} \mid \{z_{\mathrm{new},c}^{(g,t)}\}, \bm\omega_{\mathrm{new}}^{(g,t)} \right) = - \frac{1}{2} \left(\mathbf V_{\mathrm{new}}^{(g,t)}\right)^\top \bm\Lambda_{\mathrm{new}}^{(g,t)} \mathbf V_{\mathrm{new}}^{(g,t)} + \left(\mathbf V_{\mathrm{new}}^{(g,t)}\right)^\top \bm\eta_{\mathrm{new}}^{(g,t)} + C,
\]
where \(C\) is constant in \(\mathbf V_{\mathrm{new}}^{(g,t)}\), and
\begin{align}
    \bm\Lambda_{\mathrm{new}}^{(g,t)} &= \left( \bm\Sigma_V^{(g,t)} \right)^{-1} + \bm\Omega_{\mathrm{new}}^{(g,t)},
    \label{eq:supp_Vnew_block_precision}
    \\
    \bm\eta_{\mathrm{new}}^{(g,t)} &= \left( \bm\Sigma_V^{(g,t)} \right)^{-1} \bm\mu_V^{(g,t)} + \bm\kappa_{\mathrm{new}}^{(g,t)}.
    \label{eq:supp_Vnew_block_linear}
\end{align}
Completing the square yields the Gaussian update
\begin{equation}
\label{eq:supp_Vnew_block_conditional}
    \mathbf V_{\mathrm{new}}^{(g,t)} \mid \{z_{\mathrm{new},c}^{(g,t)}\}, \bm\omega_{\mathrm{new}}^{(g,t)} \sim \mathcal N_{J_g-1} \left( \left[\bm\Lambda_{\mathrm{new}}^{(g,t)}\right]^{-1} \bm\eta_{\mathrm{new}}^{(g,t)}, \left[\bm\Lambda_{\mathrm{new}}^{(g,t)}\right]^{-1} \right).
\end{equation}
The draw from \eqref{eq:supp_Vnew_block_conditional} is used as \(\mathbf V_{\mathrm{new}}^{(g,t)}\) in the response prediction step.

For response prediction, the stacked new predictor coordinate is formed as
\[
    \mathbf V_{\mathrm{new}}^{(t)} = \left[ \left(\mathbf V_{\mathrm{new},1}^{(t)}\right)^\top,\ldots, \left(\mathbf V_{\mathrm{new},G}^{(t)}\right)^\top \right]^\top,
\]
where
\[
    \mathbf V_{\mathrm{new},g}^{(t)} =
    \begin{cases}
        \mathbf V_{\mathrm{new}}^{(g,t)},& \gamma_g^{(t)}=1,\\
        \mathbf 0_{J_g-1},& \gamma_g^{(t)}=0.
    \end{cases}
\]
The zero padding for inactive groups has no effect on the regression prediction because the corresponding rows of \(\mathbf B^{(t)}\) are zero.

Conditional on \(\mathbf V_{\mathrm{new}}^{(t)}\), the response coordinate is generated from the regression layer. From \eqref{eq:supp_alpha_conditional}, the intercept conditional on the training draw satisfies
\[
    \boldsymbol\alpha^{(t)} \mid \mathbf U,\mathbf V, \mathbf B^{(t)},\bm\Sigma_U^{(t)},\bm\gamma^{(t)} \sim \mathcal N_q \left( \bar{\mathbf U}^{(t)} - \left(\mathbf B^{(t)}\right)^\top \bar{\mathbf V}^{(t)}, \frac{1}{N}\bm\Sigma_U^{(t)} \right).
\]
For a new response coordinate,
\[
    \mathbf U_{\mathrm{new}}^{(t)} = \boldsymbol\alpha^{(t)} + \left(\mathbf B^{(t)}\right)^\top \mathbf V_{\mathrm{new}}^{(t)} + \boldsymbol\varepsilon_{\mathrm{new}}^{(t)}, \qquad \boldsymbol\varepsilon_{\mathrm{new}}^{(t)} \sim \mathcal N_q(\mathbf 0,\bm\Sigma_U^{(t)}).
\]
Integrating over the Gaussian conditional distribution of \(\boldsymbol\alpha^{(t)}\) gives the single-subject marginal predictive distribution
\begin{equation}
\label{eq:supp_prediction_U}
    \mathbf U_{\mathrm{new}}^{(t)} \mid \mathbf V_{\mathrm{new}}^{(t)},\mathcal D \sim \mathcal N_q \left[ \bar{\mathbf U}^{(t)} + \left(\mathbf B^{(t)}\right)^\top \left\{ \mathbf V_{\mathrm{new}}^{(t)} - \bar{\mathbf V}^{(t)} \right\}, \left(1+\frac{1}{N}\right) \bm\Sigma_U^{(t)} \right].
\end{equation}

Because the inverse ALR map is nonlinear, we approximate the response mixture weights for posterior draw \(t\) by Monte Carlo averaging. Draw
\[
    \mathbf U_{\mathrm{new}}^{(t,\ell)} \sim \mathcal N_q \left[ \bar{\mathbf U}^{(t)} + \left(\mathbf B^{(t)}\right)^\top \left\{ \mathbf V_{\mathrm{new}}^{(t)} - \bar{\mathbf V}^{(t)} \right\}, \left(1+\frac{1}{N}\right) \bm\Sigma_U^{(t)} \right], \qquad \ell=1,\ldots,M_{\mathrm{pred}},
\]
and set
\[
\widehat{\bm\pi}_{\mathrm{new}}^{Y(t)}
=
\frac{1}{M_{\mathrm{pred}}}
\sum_{\ell=1}^{M_{\mathrm{pred}}}
\operatorname{ALR}^{-1}_K
\left(\mathbf U_{\mathrm{new}}^{(t,\ell)}\right).
\]
The posterior predictive response density for this retained draw is then
\begin{equation}
\label{eq:supp_prediction_density}
    \widehat p_{\mathrm{new}}^{(t)}(\mathbf y) = \sum_{k=1}^{K} \widehat\pi_{\mathrm{new},k}^{Y(t)} \, \mathcal N_{d_Y} \left( \mathbf y \mid \bm\mu_k^{Y(t)}, \bm\Sigma_k^{Y(t)} \right).
\end{equation}
Posterior predictive summaries are obtained by averaging \eqref{eq:supp_prediction_density} over retained posterior draws.

\section{Simulation Design and Additional Results}
\label{supp:simulation}

\subsection{Data Generation}
\label{sec:simulation_dgp}

For each simulation replicate, we generate data for $N = 400$ subjects, evenly split into training and test sets of $200$ each. We first sample latent predictor coordinates $\mathbf{V}_i^{(g)} \in \mathbb{R}^{J_g - 1}$ for each predictor $g = 1, \ldots, G$ from $\mathcal{N}(\mathbf{0}, \bm{\Sigma}_V^{(g)})$, where $\bm{\Sigma}_V^{(g)}$ is diagonal with entries drawn uniformly from $[0.5, 1]$.

The coefficient matrix $\mathbf{B}$ has group-sparse structure with $G_{\mathrm{active}} = 2$ active predictors; nonzero entries are drawn uniformly from $\pm[0.2, 0.5]$. The residual covariance $\bm{\Sigma}_U$ is diagonal with entries drawn uniformly from $[0.5, 1.5]$. We set $\boldsymbol\alpha=\mathbf0$ and generate the response coordinates according to
\[
\mathbf U_i=\mathbf B^\top\mathbf V_i+\bm\varepsilon_i, \qquad \bm\varepsilon_i\sim\mathcal N(\mathbf0,\bm\Sigma_U).
\]
The first two predictor groups are active, whereas all remaining groups have zero coefficient blocks.

Response mixing weights are $\bm{\pi}_i^Y = \mathrm{ALR}_K^{-1}(\mathbf{U}_i)$. The shared component means follow a gap structure $\bm{\mu}_k^Y = \bm{\mu}_{\mathrm{base}} + (k-1)\Delta \mathbf{1}$, where $\bm{\mu}_{\mathrm{base}}$ is a base vector, and all component covariances share an AR(1) structure $\bm{\Sigma}_k^Y = \mathbf{D}_\sigma \mathbf{R}_\rho \mathbf{D}_\sigma$ with $(\mathbf{R}_\rho)_{ij} = \rho^{|i-j|}$ and $\mathbf{D}_\sigma = \mathrm{diag}(\sigma_1, \ldots, \sigma_d)$. We set $\Delta = 2$, $\rho = 0.1$, and $\sigma_l = \sqrt{0.6}$ for all components. Predictor distributions follow analogous specifications with $J_g$ components. Each subject contributes a common sample size $C$ per distribution in the simulations, that is, $N_i^Y=N_i^{(g)}=C$ for all $i$ and $g$, where $C$ varies by scenario.

\subsection{Scenario 3: Comparison with Univariate Methods}
\label{sec:sim_univariate}

Scenario 3 assesses the benefit of native multivariate modeling by increasing the dimension to $d_Y=d_g=5$ and comparing \spider{} with DRN and Projected Reg. We set $G=10$, $G_{\mathrm{active}}=2$, and $C=50$, and vary $K_0=J_0\in\{3,4,5\}$. Because DRN and Projected Reg. operate only on univariate distributions and do not perform group variable selection, they are supplied with the two oracle-active predictors. For each such multivariate predictor $\mathbf X^{(g)}$, we extract its first principal component via PCA to obtain a univariate predictor distribution. On the response side, we fit a separate univariate regression for each marginal dimension $Y_1,\ldots,Y_{d_Y}$ and reconstruct the multivariate predictive distribution under the independence approximation
\[
    \widehat p(\mathbf Y)=\prod_{l=1}^{d_Y}\widehat p(Y_l).
\]
This adaptation discards cross-dimensional dependence in both predictors and responses, whereas \spider{} models the full multivariate distributions directly and performs selection over all ten candidate predictors. Results are reported in Table~\ref{tab:sim_univariate}.

\begin{table}[htbp]
\centering
\caption{Scenario 3: comparison with univariate distributional regression methods.}
\label{tab:sim_univariate}
\begingroup
\small
\setlength{\tabcolsep}{6pt}
\renewcommand{\arraystretch}{1.02}
\begin{tabular}{@{}ccccccc@{}}
\toprule
& \multicolumn{3}{c}{SWD (SE)}
& \multicolumn{3}{c}{\spider{} Variable Selection} \\
\cmidrule(lr){2-4} \cmidrule(lr){5-7}
$K_0=J_0$ & \spider{} & DRN & Projected Reg.
& MCC & TPR & FPR \\
\midrule
3 & \textbf{0.685} (0.006) & 1.114 (0.005) & 0.965 (0.003) & 0.867 & 1.000 & 0.078 \\
4 & \textbf{0.982} (0.007) & 1.615 (0.006) & 1.384 (0.004) & 0.967 & 1.000 & 0.018 \\
5 & \textbf{1.288} (0.009) & 2.108 (0.008) & 1.793 (0.005) & 0.955 & 1.000 & 0.026 \\
\bottomrule
\end{tabular}
\endgroup
\end{table}

\spider{} substantially outperforms both univariate baselines across all settings. At $K_0=J_0=3$, its SWD of $0.685$ improves on Projected Reg. and DRN by $29.0\%$ and $38.5\%$, respectively. At $K_0=J_0=5$, the corresponding improvements are $28.2\%$ and $38.9\%$. These gains indicate that reducing each multivariate predictor to a single principal component and reconstructing the response under marginal independence can discard dependence information that is important for distribution-to-distribution prediction. Variable selection remains stable, with MCC exceeding $0.86$, TPR equal to $1.000$, and FPR below $0.08$ throughout.

\section{Application Details}
\label{supp:application_details}

This section contains the ligand and receptor gene table used to define the cell-cell communication analysis in the OneK1K data application.

\begin{table}[htbp]
\centering
\caption{Ligand and receptor genes used in the analysis, derived from significant CellChat L-R pairs ($p < 0.05$).}
\label{tab:supp_LR_genes}
\begingroup
\small
\setlength{\tabcolsep}{4pt}
\renewcommand{\arraystretch}{1.05}
\begin{tabularx}{\textwidth}{
  >{\centering\arraybackslash}p{2.2cm}
  >{\raggedright\arraybackslash}X
  >{\centering\arraybackslash}p{1.6cm}
  >{\raggedright\arraybackslash}X
}
\toprule
\textbf{Senders} & \textbf{Ligands} &
\textbf{Target} & \textbf{Receptors} \\
\midrule
B, CD4$^+$ T, Mono, NK & CD99, HLA-A, HLA-B, HLA-C, HLA-E, HLA-F, LCK, LGALS9, MIF, PPIA & CD8$^+$ T & BSG, CD44, CD74, CD8A, CD8B, CD99, CXCR4, KLRK1, P4HB, PTPRC \\
\midrule
B, CD4$^+$ T, CD8$^+$ T, Mono & CD55, CD69, CD99, CLEC2B, CLEC2D, HLA-E, LGALS9, MIF, PPIA & NK & ADGRE5, BSG, CD44, CD74, CD94, CD99, KLRB1, KLRC1, KLRK1, NKG2A, P4HB, PTPRC \\
\midrule
CD4$^+$ T, CD8$^+$ T, Mono, NK & CD160, CD99, LGALS9, MIF, PTPRC, SELPLG, TNFSF13B & B & CD22, CD44, CD74, CD99, CXCR4, PTPRC, SELL, TNFRSF13C, TNFRSF14 \\
\midrule
CD8$^+$ T, Mono, NK & CD160, CD99, LGALS9, SELPLG & CD4$^+$ T & CD44, CD99, P4HB, PTPRC, SELL, TNFRSF14 \\
\midrule
B, CD4$^+$ T, CD8$^+$ T, NK & ANXA1, CD160, CD99, HLA-DMA, HLA-DMB, HLA-DOB, HLA-DPA1, HLA-DPB1, HLA-DQA1, HLA-DQB1, HLA-DRA, HLA-DRB1, HLA-F, MIF, PPIA & Mono & BSG, CD4, CD44, CD74, CD99, FPR1, LILRB1, PILRA, TNFRSF14 \\
\bottomrule
\end{tabularx}
\endgroup
\end{table}

\section{MCMC Diagnostics}
\label{supp:mcmc_diagnostics}

This appendix documents the MCMC convergence diagnostics for the real data analysis in Section~\ref{sec:application}. Because the \spider{} sampler includes discrete sender-inclusion indicators $\bm{\gamma}$, the posterior may concentrate on several distinct model configurations during a single run. Standard single-chain diagnostics can be misleading in this setting, as apparent jumps may simply reflect transitions between well-mixed inclusion patterns rather than poor convergence. We therefore adopt a two-pronged diagnostic approach: (i) joint inspection of the log-likelihood trajectory alongside the sampled $\bm{\gamma}$ trajectory, and (ii) conditional autocorrelation functions (ACFs) computed after stratifying the post-burn-in samples by the visited $\bm{\gamma}$ pattern, so that within-pattern mixing is assessed separately from movement between model states.

\begin{figure}[htbp]
\centering
\includegraphics[width=0.50\textwidth,height=0.33\textheight,keepaspectratio]{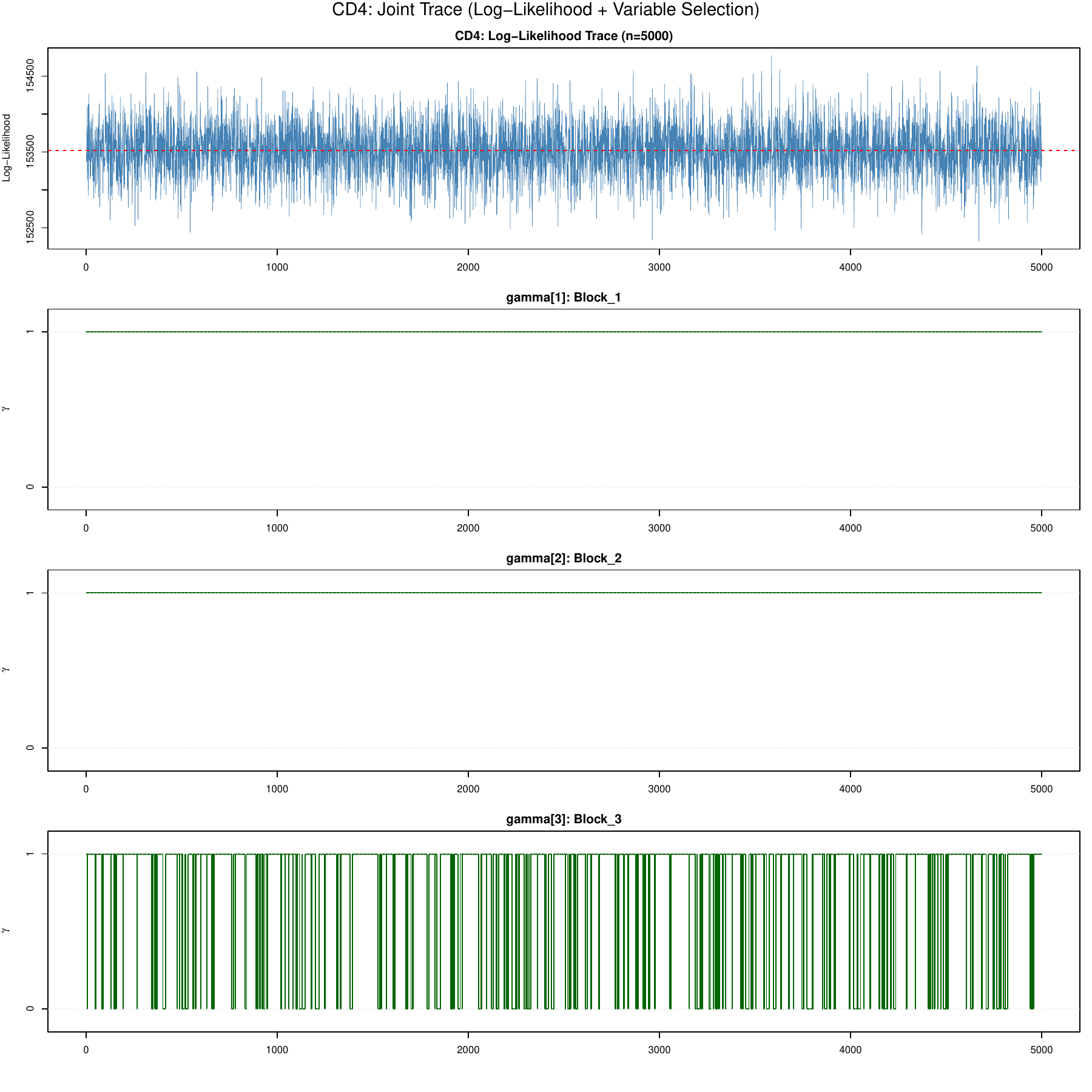}
\hfill
\includegraphics[width=0.45\textwidth,height=0.33\textheight,keepaspectratio]{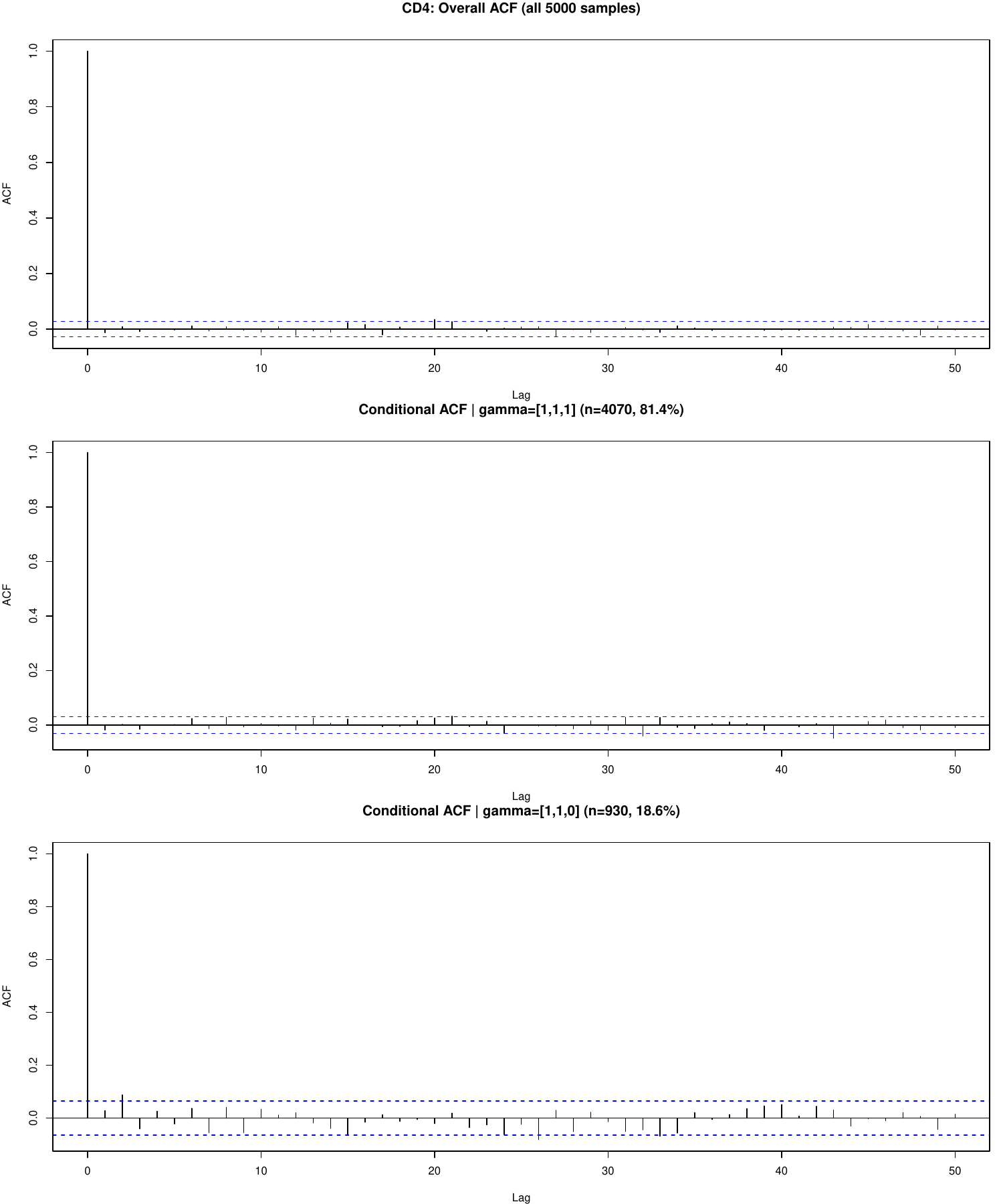}
\caption{MCMC diagnostics for CD4$^+$ T cells as target.
\emph{Left}: joint trace of the post-burn-in log-likelihood and sender-inclusion
indicators~$\bm{\gamma}$.
\emph{Right}: conditional autocorrelation function of the log-likelihood,
stratified by the visited $\bm{\gamma}$ pattern.}
\label{fig:mcmc_diag_cd4}
\end{figure}

\begin{figure}[htbp]
\centering
\includegraphics[width=0.50\textwidth,height=0.33\textheight,keepaspectratio]{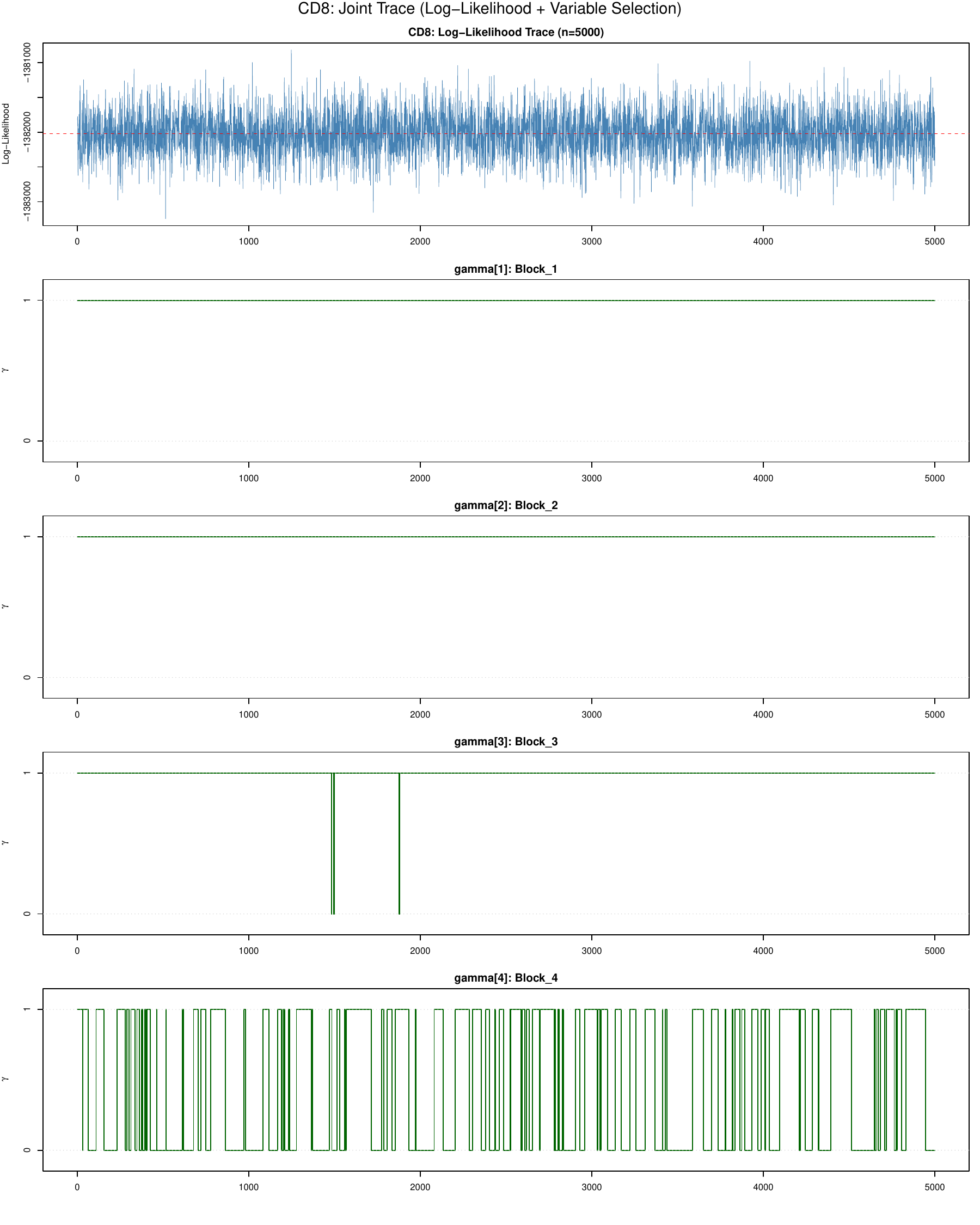}
\hfill
\includegraphics[width=0.45\textwidth,height=0.33\textheight,keepaspectratio]{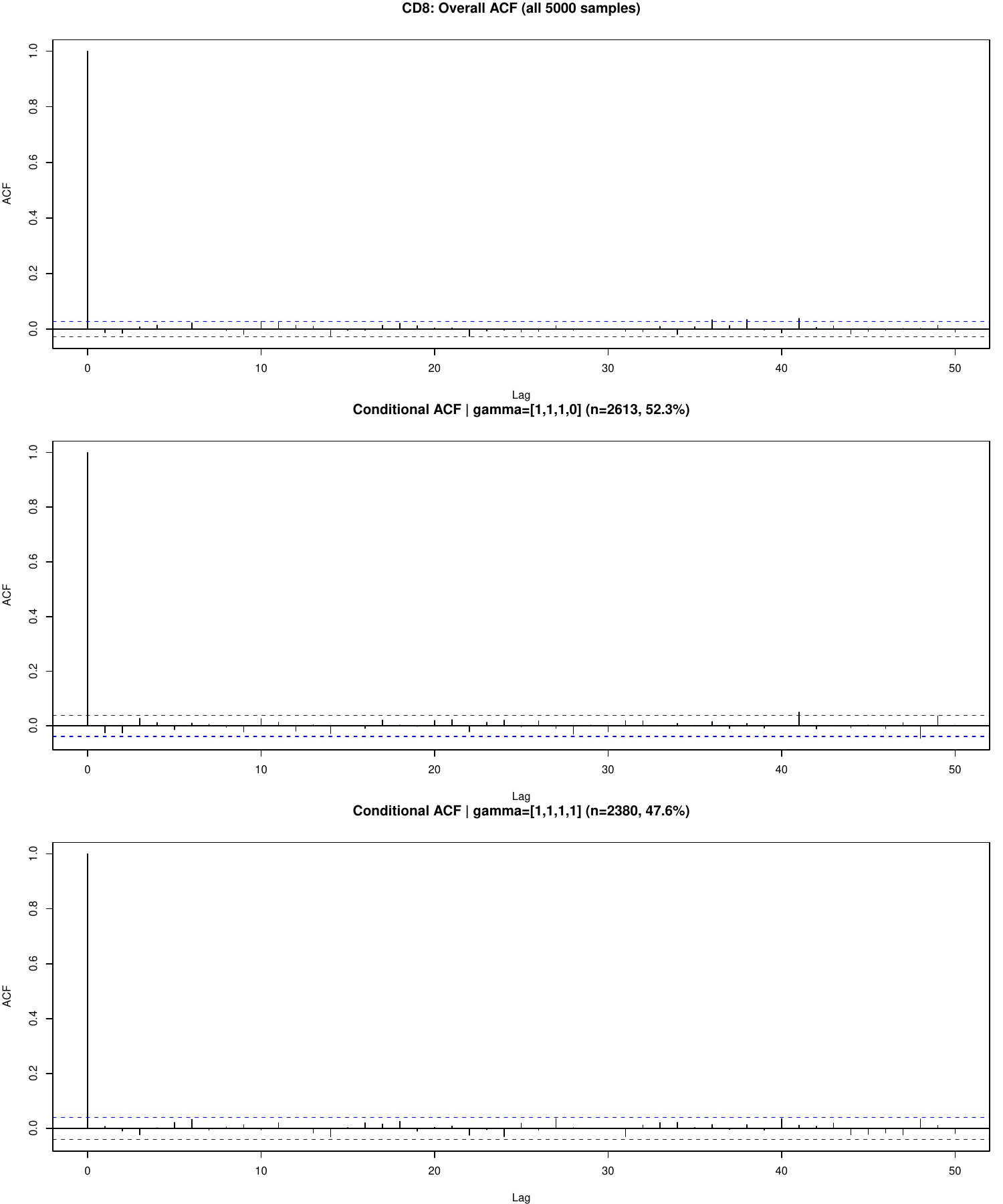}
\caption{MCMC diagnostics for CD8$^+$ T cells as target.
\emph{Left}: joint trace of the post-burn-in log-likelihood and sender-inclusion
indicators~$\bm{\gamma}$.
\emph{Right}: conditional autocorrelation function of the log-likelihood,
stratified by the visited $\bm{\gamma}$ pattern.}
\label{fig:mcmc_diag_cd8}
\end{figure}

\begin{figure}[htbp]
\centering
\includegraphics[width=0.50\textwidth,height=0.33\textheight,keepaspectratio]{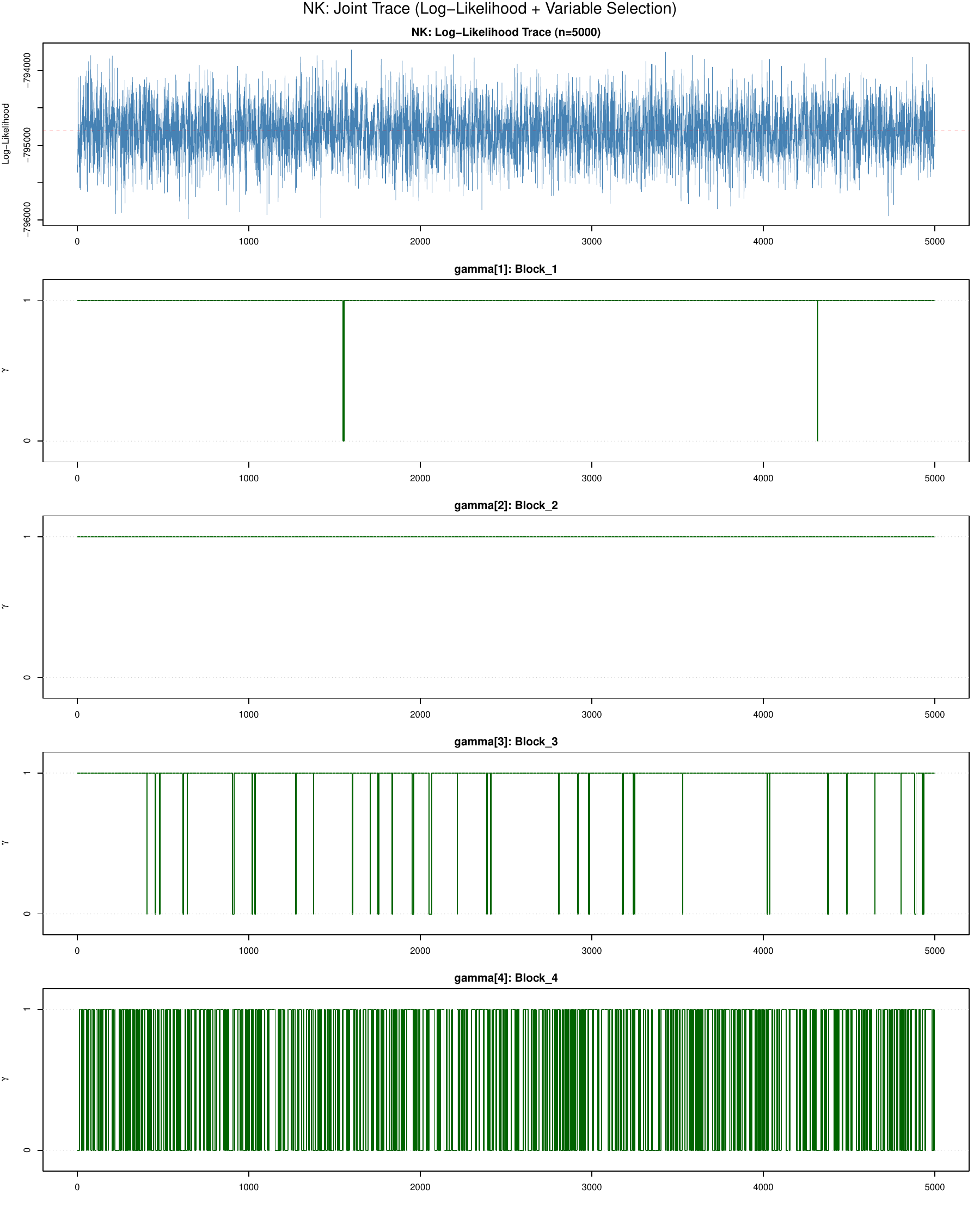}
\hfill
\includegraphics[width=0.45\textwidth,height=0.33\textheight,keepaspectratio]{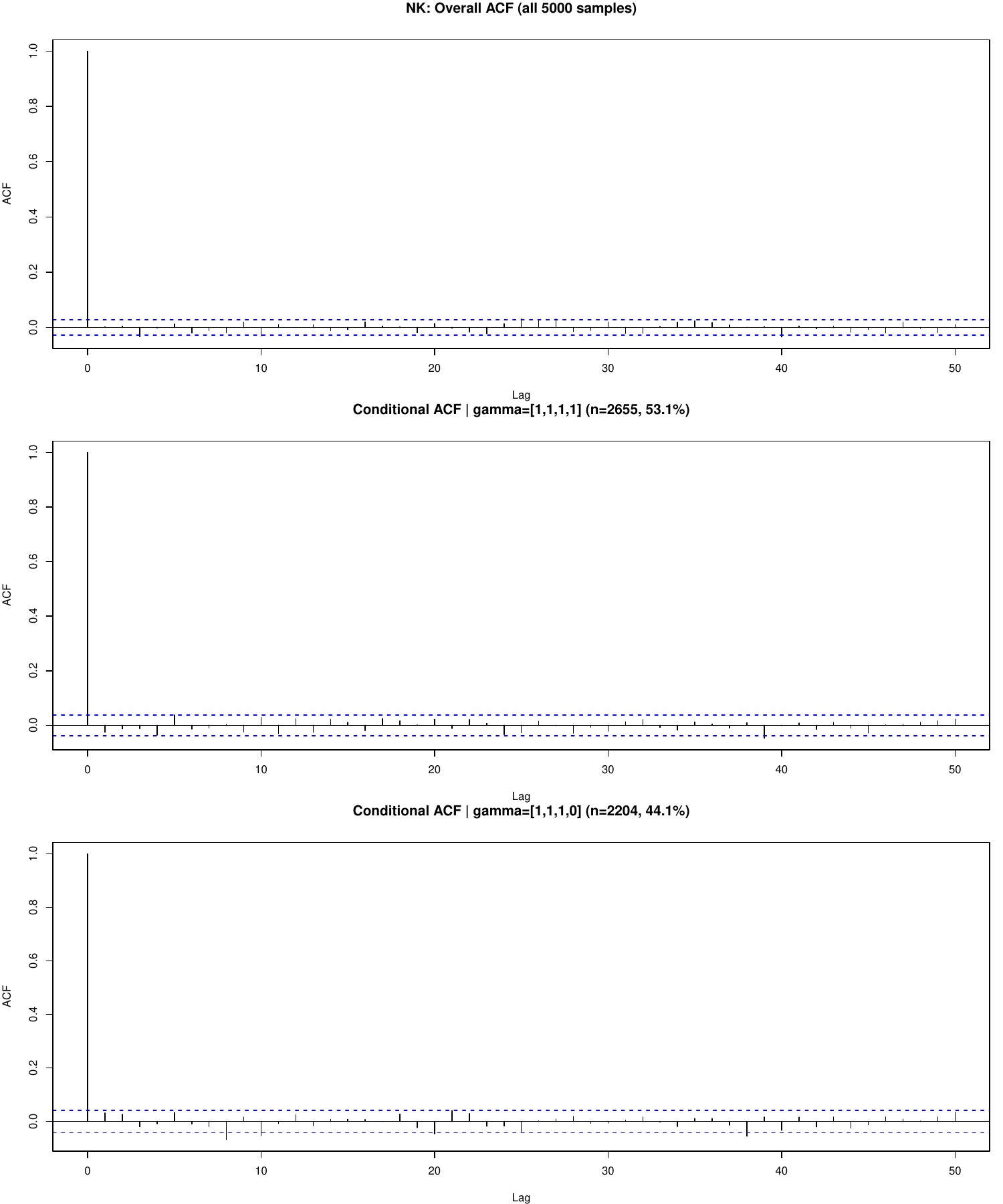}
\caption{MCMC diagnostics for NK cells as target.
\emph{Left}: joint trace of the post-burn-in log-likelihood and sender-inclusion
indicators~$\bm{\gamma}$.
\emph{Right}: conditional autocorrelation function of the log-likelihood,
stratified by the visited $\bm{\gamma}$ pattern.}
\label{fig:mcmc_diag_nk}
\end{figure}

\begin{figure}[htbp]
\centering
\includegraphics[width=0.50\textwidth,height=0.33\textheight,keepaspectratio]{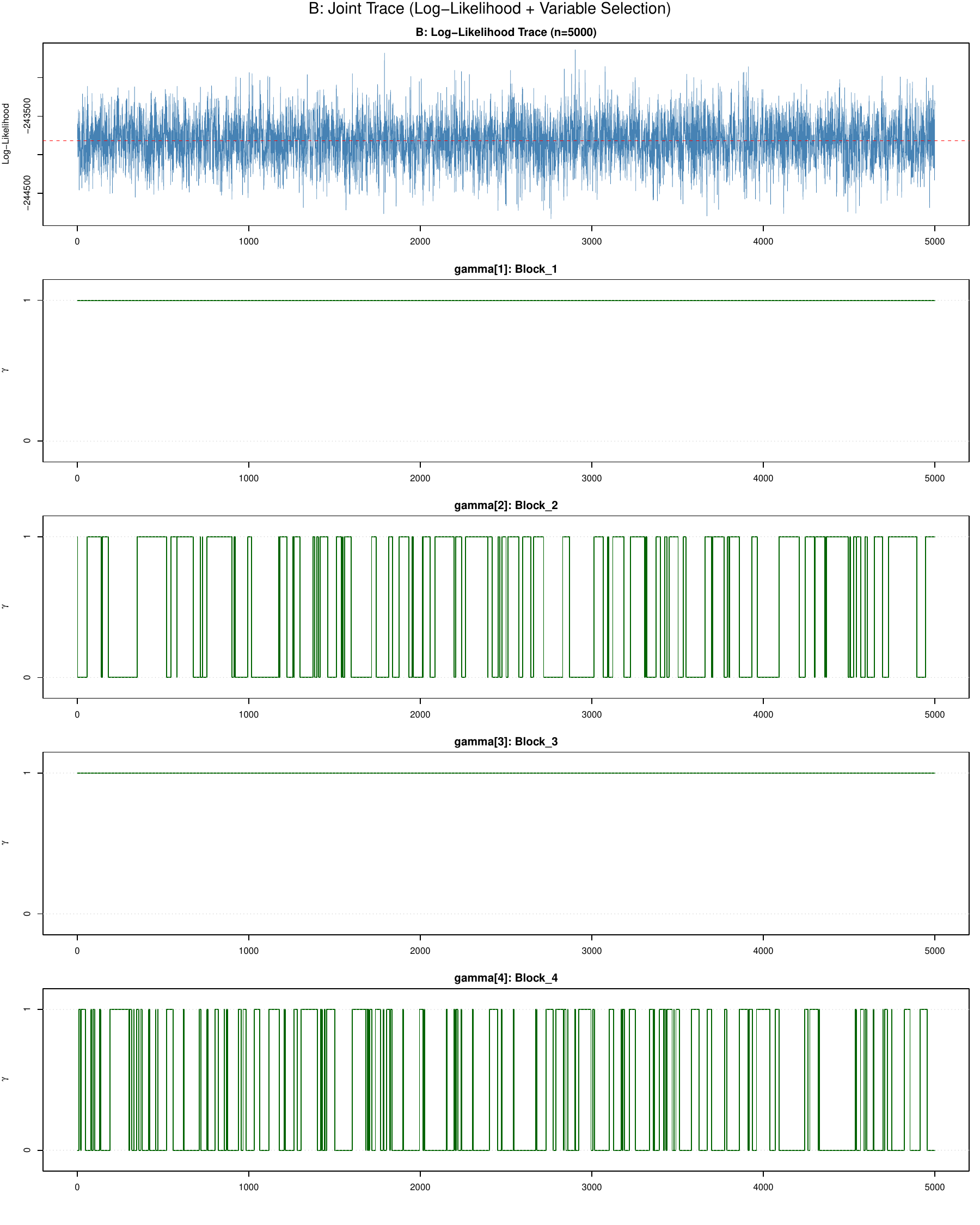}
\hfill
\includegraphics[width=0.45\textwidth,height=0.33\textheight,keepaspectratio]{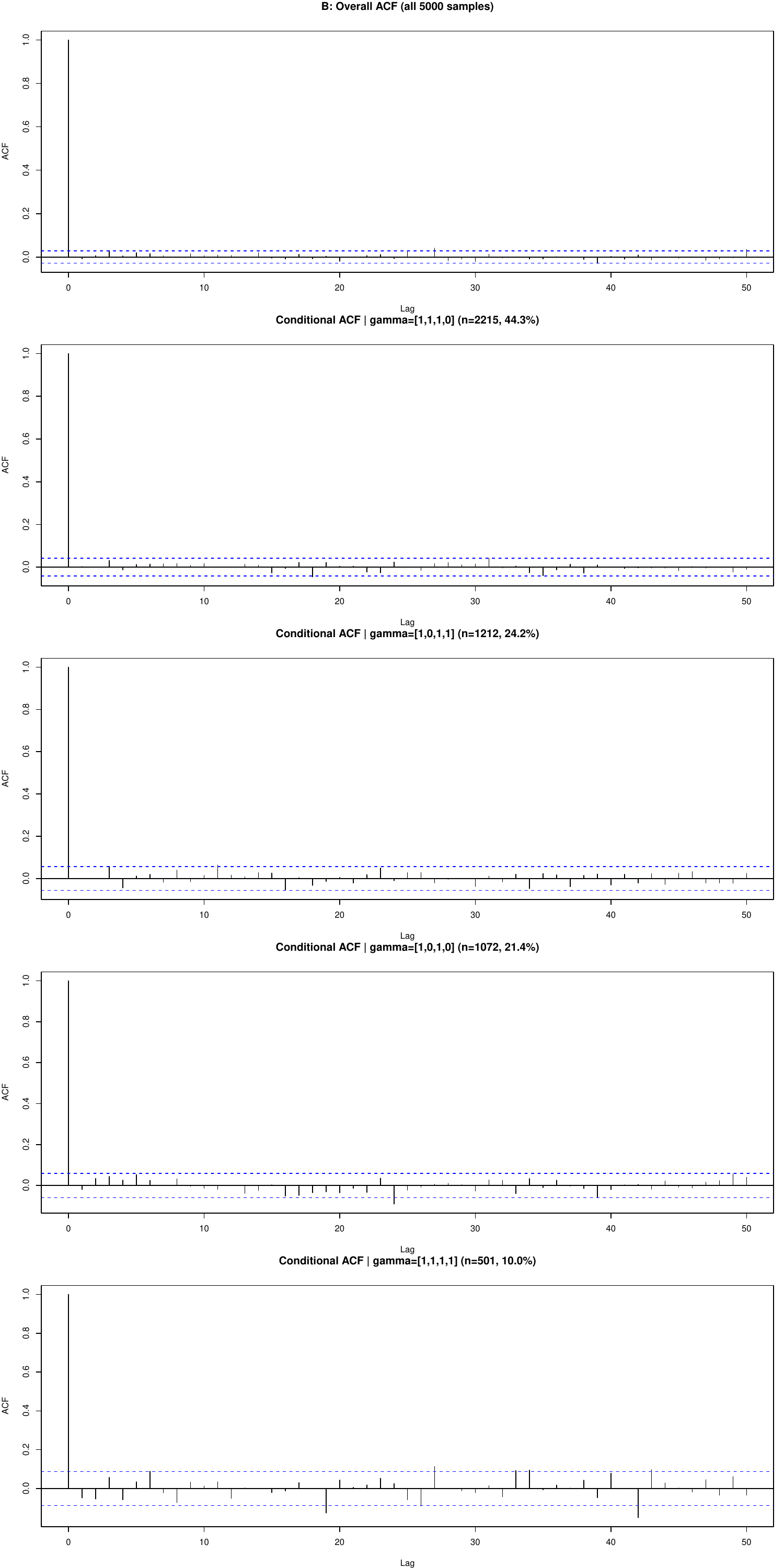}
\caption{MCMC diagnostics for B cells as target.
\emph{Left}: joint trace of the post-burn-in log-likelihood and sender-inclusion
indicators~$\bm{\gamma}$.
\emph{Right}: conditional autocorrelation function of the log-likelihood,
stratified by the visited $\bm{\gamma}$ pattern.}
\label{fig:mcmc_diag_b}
\end{figure}

\begin{figure}[htbp]
\centering
\includegraphics[width=0.50\textwidth,height=0.33\textheight,keepaspectratio]{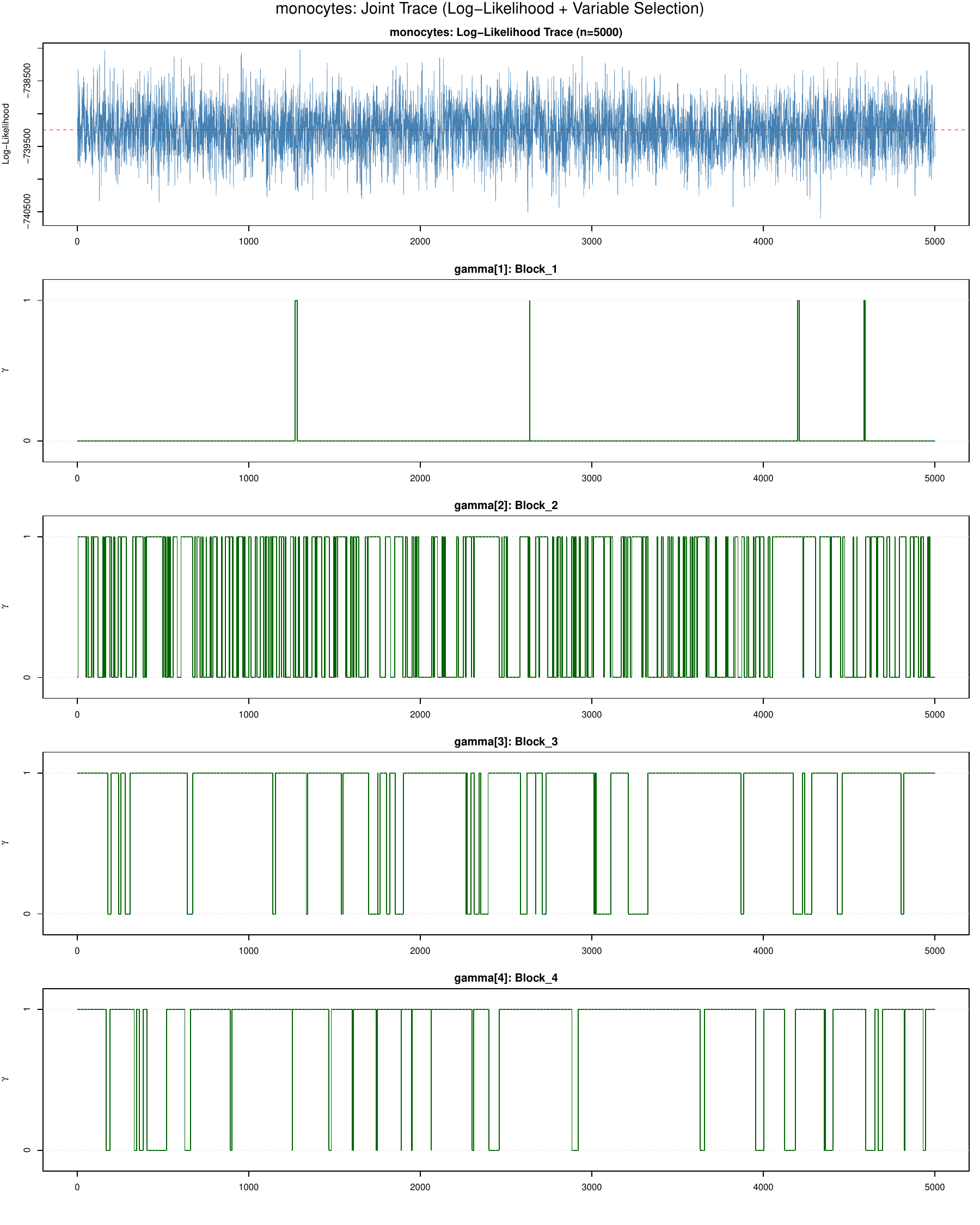}
\hfill
\includegraphics[width=0.45\textwidth,height=0.33\textheight,keepaspectratio]{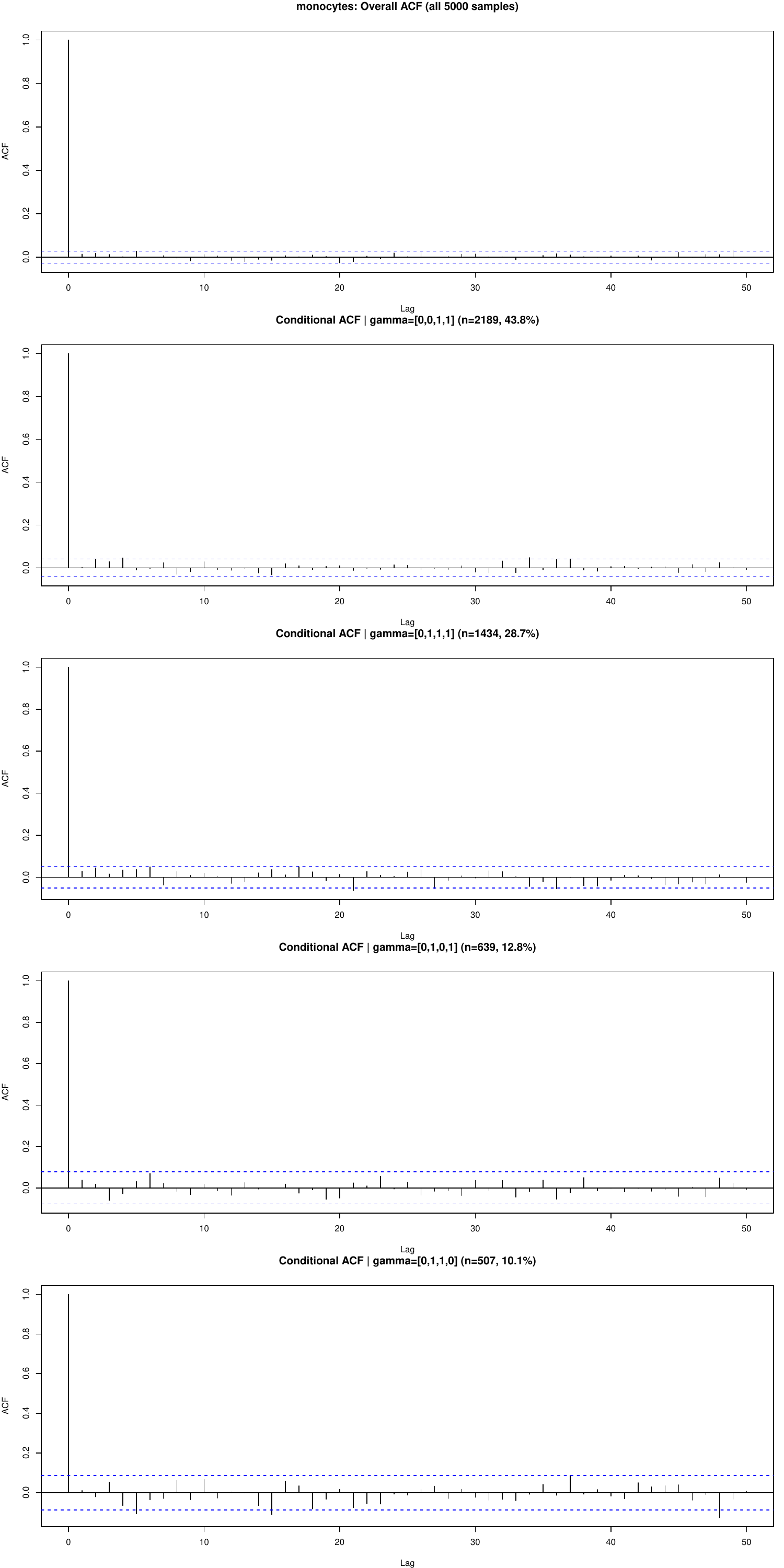}
\caption{MCMC diagnostics for monocytes as target.
\emph{Left}: joint trace of the post-burn-in log-likelihood and sender-inclusion
indicators~$\bm{\gamma}$.
\emph{Right}: conditional autocorrelation function of the log-likelihood,
stratified by the visited $\bm{\gamma}$ pattern.}
\label{fig:mcmc_diag_mono}
\end{figure}

For each target cell type, we refit the WAIC-selected model using $10{,}000$ MCMC iterations with $5{,}000$ burn-in iterations, yielding $5{,}000$ post-burn-in samples. The left panels of Figures~\ref{fig:mcmc_diag_cd4}--\ref{fig:mcmc_diag_mono} display the post-burn-in log-likelihood together with the sampled sender-inclusion indicators $\bm{\gamma}$. The right panels show the conditional ACFs, computed after stratifying the post-burn-in samples by the visited $\bm{\gamma}$ patterns; this allows within-pattern mixing to be inspected independently of the inter-model transitions.

Across all five target cell types, the log-likelihood traces exhibit rapid stabilization after burn-in, with no apparent trends or transient behavior within any visited $\bm{\gamma}$ configuration. The conditional ACFs decay to near zero without persistent long-lag autocorrelation for the dominant visited patterns, indicating adequate within-pattern mixing.

\end{document}